%% file: paper-recurrence.tex
\pdfoutput=1
\documentclass{sig-alternate}
\usepackage[usenames,dvipsnames]{color}
\usepackage{multirow}
\usepackage{booktabs}
\newcommand{\ra}[1]{\renewcommand{\arraystretch}{#1}}
\usepackage{caption}
\usepackage{subcaption}
\usepackage{tabularx}
\usepackage[inline]{enumitem}
\usepackage{etoolbox}
\usepackage{times}
\usepackage{balance}
\newif\ifcameraready

\camerareadytrue

\ifcameraready
\newcommand\todo[1]{}
\newcommand{\lada}[1]{}
\newcommand{\jon}[1]{}
\newcommand{\jure}[1]{}
\else
\newcommand\todo{\textcolor{red}}
\newcommand{\lada}[1]{{\color{magenta}[LADA: #1]}}
\newcommand{\jon}[1]{{\color{blue}[JON: #1]}}
\newcommand{\jure}[1]{{\color{green}[CRISTIAN: #1]}}
\fi

\newcommand{\hide}[1]{}
\newcommand{\xhdr}[1]{\vspace{1mm}\noindent{{\bf #1.}}}
\newcommand{\xhdrr}[1]{\vspace{1mm}\noindent{{\bf #1}}}

\newcommand\deem{\textcolor{Gray}}
\providecommand{\e}[1]{\ensuremath{\times}10\textsuperscript{#1}}
\newcommand\pvallow{\textit{p}\textless10\textsuperscript{-10}}

\graphicspath{{./figures/}}

\newcommand{\IconIfTable}{\includegraphics[scale=0.5]{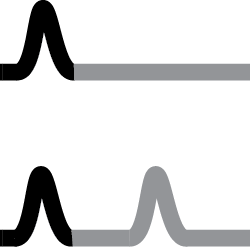}}
\newcommand{\IconSizeTable}{\includegraphics[scale=0.5]{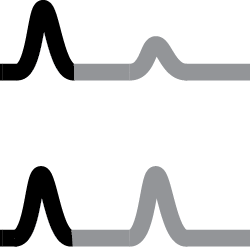}}
\newcommand{\IconWhenTable}{\includegraphics[scale=0.5]{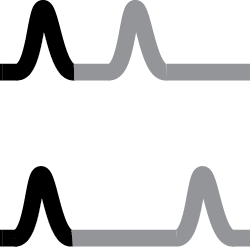}}
\newcommand{\IconIf}{\raisebox{-0.5ex}{\includegraphics[scale=0.45]{icon_if}}}
\newcommand{\IconSize}{\raisebox{-0.5ex}{\includegraphics[scale=0.45]{icon_size}}}
\newcommand{\IconWhen}{\raisebox{-0.5ex}{\includegraphics[scale=0.45]{icon_when}}}
\robustify{\IconIf}
\robustify{\IconSize}
\robustify{\IconWhen}

\permission{}
\conferenceinfo{}{}
\copyrightetc{}
\crdata{}

\begin{document}

\title{Do Cascades Recur?}

\numberofauthors{1}
\author{Justin Cheng$^1$, Lada A Adamic$^2$, Jon Kleinberg$^3$, Jure Leskovec$^4$\\
\affaddr{$^{1,4}$Stanford University, $^2$Facebook, $^3$Cornell University}\\
\email{$^{1,4}$\{jcccf, jure\}@cs.stanford.edu, $^2$ladamic@fb.com, $^3$kleinber@cs.cornell.edu}
}

\maketitle

\begin{abstract}
\input{000abstract}
\end{abstract}

\vspace{1mm}


\vspace{0mm} 

\section{Introduction}
\label{sec:intro}
\input{010intro}

\section{Technical Preliminaries}
\label{sec:recurrence}
\input{020data}

\section{Characterizing Recurrence}
\label{sec:characterizing}
\input{030characteristics1}
\input{035characteristics2}

\section{Modeling Recurrence}
\label{sec:model}
\input{040model}
\label{sec:simulation}
\input{045simulation}

\section{Predicting Recurrence}
\label{sec:predicting}
\input{050prediction}

\section{Related Work}
\label{sec:related}
\input{060related}

\section{Discussion and Conclusion}
\label{sec:conclusion}
\input{070conclusion}

\bibliographystyle{abbrv}
\bibliography{refs}

\end{document}

%% file: 000abstract.tex

Cascades of information-sharing are a primary mechanism by which content reaches its audience on social media, and an active line of research has studied how such cascades, which form as content is reshared from person to person, develop and subside.
In this paper, we perform a large-scale analysis of cascades on Facebook over significantly longer time scales, and find that a more complex picture emerges, in which many large cascades {\em recur}, exhibiting multiple bursts of popularity with periods of quiescence in between.
We characterize recurrence by measuring the time elapsed between bursts, their overlap and proximity in the social network, and the diversity in the demographics of individuals participating in each peak.
We discover that content virality, as revealed by its initial popularity, is a main driver of recurrence, with the availability of multiple copies of that content helping to spark new bursts.
Still, beyond a certain popularity of content, the rate of recurrence drops as cascades start exhausting the population of interested individuals.
We reproduce these observed patterns in a simple model of content recurrence simulated on a real social network.
Using only characteristics of a cascade's initial burst, we demonstrate strong performance in predicting whether it will recur in the future.

%% file: 010intro.tex

\begin{figure}[tb]
    \centering
    \includegraphics[width=\columnwidth]{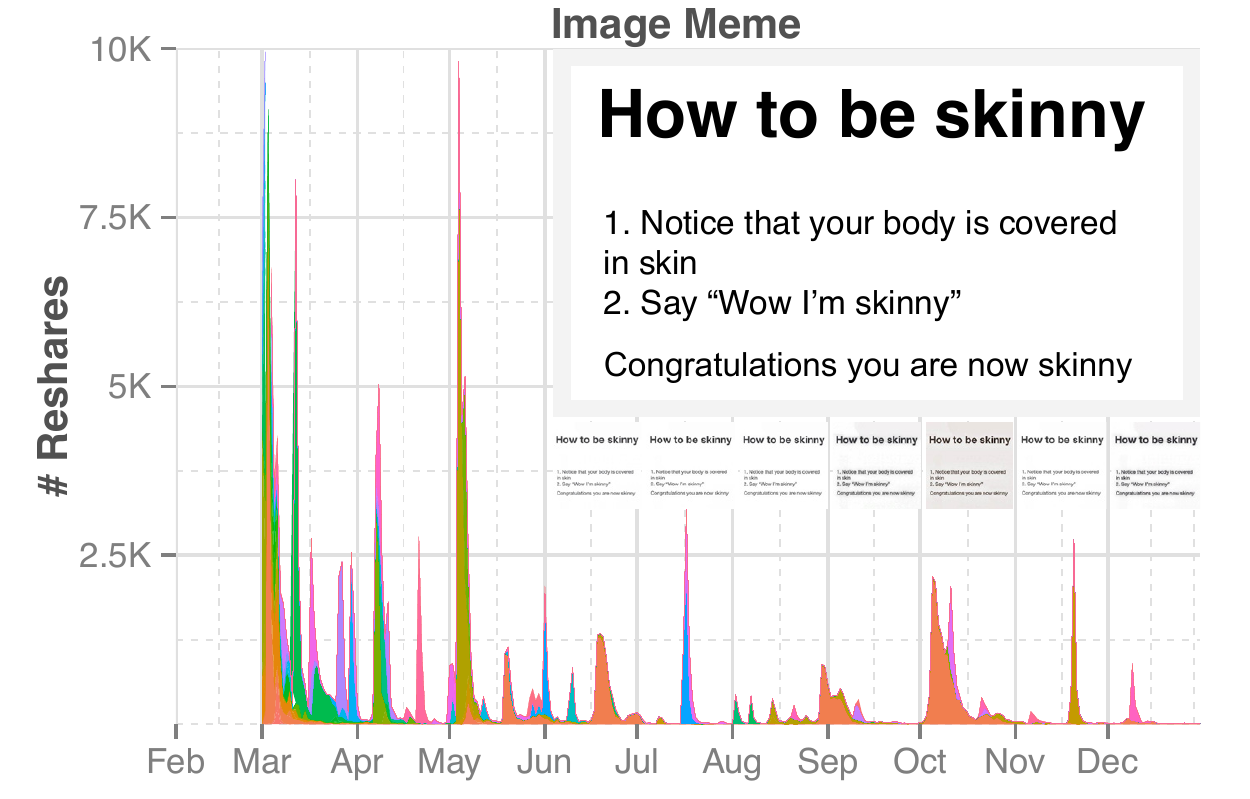}
    \caption{An example of a image meme that has recurred, or resurfaced in popularity multiple times, sometimes as a continuation of the same copy, and sometimes as a new copy of the same meme (example copies are shown as thumbnails).  This recurrence appears as multiple peaks in the plot of reshares as a function of time.}
    \label{fig:examplememe}
\end{figure}

\begin{figure*}[tb]
    \centering
    \begin{tabular}{@{}cccc@{}}
    \hspace{-0.1cm}
    \includegraphics[width=0.24\textwidth]{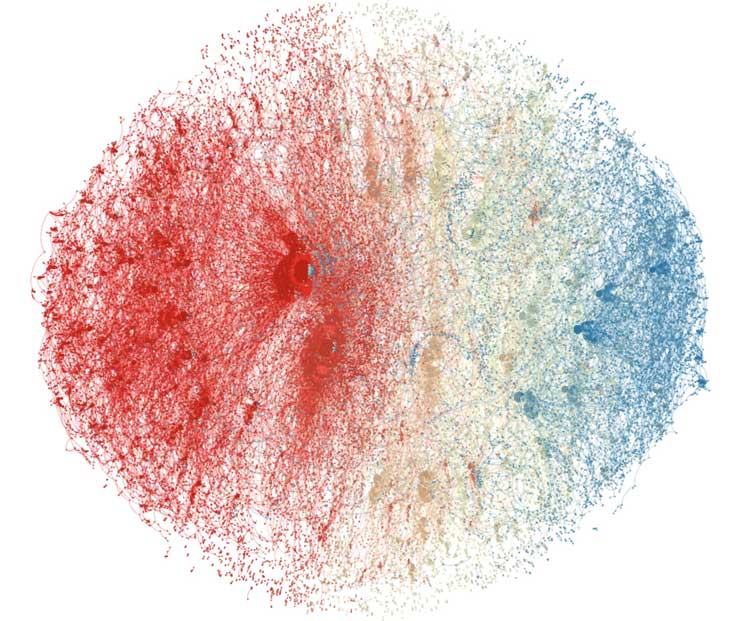}
    \hspace{-0.1cm}
    &
    \hspace{-0.1cm}
    \includegraphics[width=0.24\textwidth]{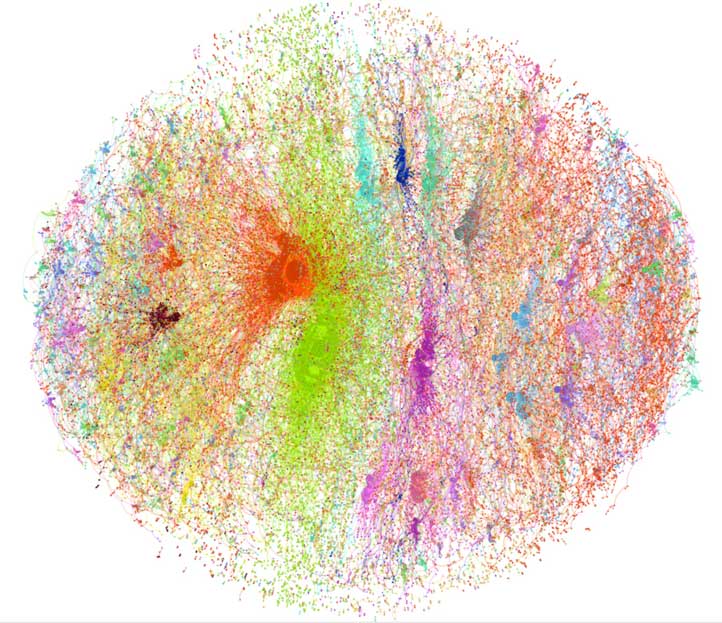}
    \hspace{-0.1cm}
    &
    \hspace{-0.1cm}
    \includegraphics[width=0.24\textwidth]{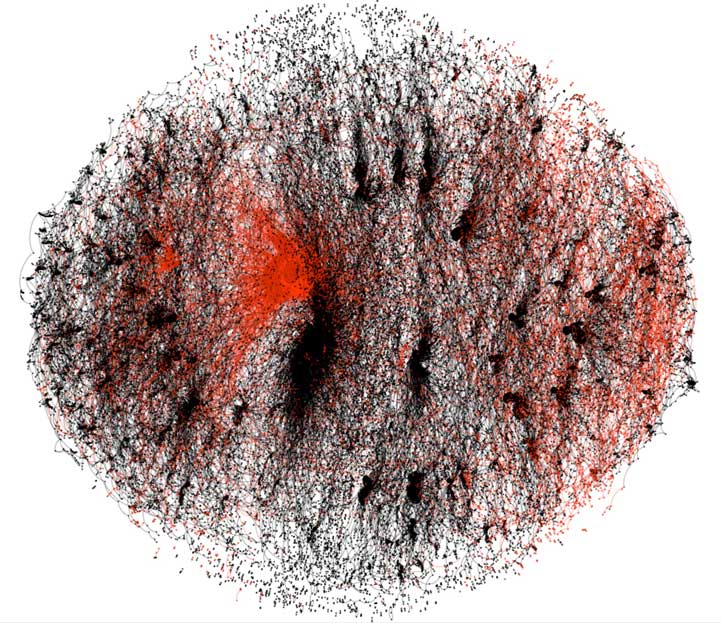}
    \hspace{-0.1cm}
    &
    \hspace{-0.1cm}
    \includegraphics[width=0.24\textwidth]{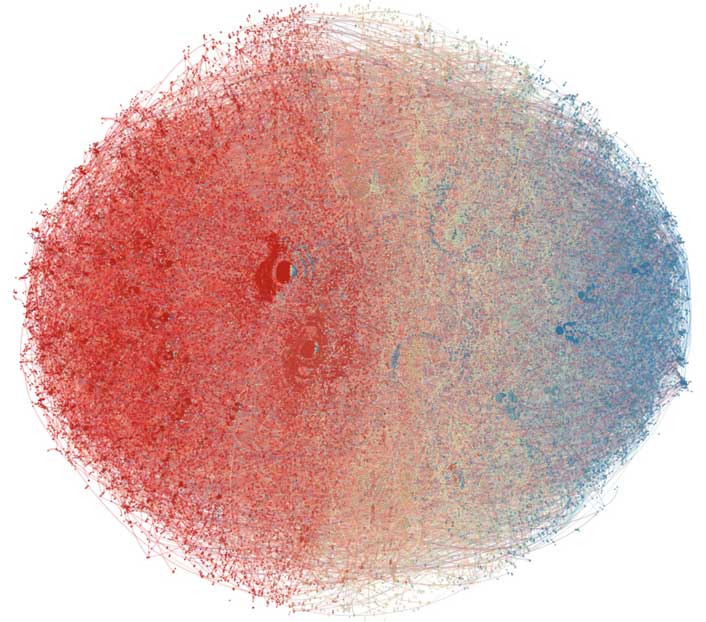}\\
    \small{(a)} & \small{(b)} & \small{(c)} & \small{(d)}
    \end{tabular}
    \caption{
    (a) The diffusion cascade of the example meme from Figure \ref{fig:examplememe} as it spreads over time, colored from red (early) to blue (late). Only reshares that prompted subsequent reshares are shown.
    (b) The cascade is made up of separately introduced copies of the same content; in this drawing of the cascade from (a), each copy is represented in a different color.
    (c) Sometimes, individual copies experience a resurgence in popularity; again we draw the cascade from (a), but now highlight a single resurgent copy in red with the spread of all other copies depicted in black.
    (d) A different network on the same set of users who took part in the cascade, showing friendship edges rather than reshare edges. These edges span reshares across copies and time, showing that multiple copies of the meme are not well-separated in the friendship network.
    }
    \label{fig:communities}
\end{figure*}

In many online social networks, people share content in the form of
photos, videos, and links with one another.  As others reshare this
content with their friends or followers in turn, cascades of resharing
can develop \cite{cheng2014can}.  Substantial previous work has
studied the formation of such information cascades with the aim of
characterizing and predicting their growth
\cite{bakshy2012role,gruhl2004information,yang2010predicting}.  
Cascades tend to be bursty, with a spike of activity occurring within a few
days of the content's introduction into the network
\cite{leskovec2007patterns,myers2014bursty}. This property
forms the backdrop to a line of temporal analyses that focus on
the basic rising-and-falling pattern that characterizes the
initial onset of a cascade
\cite{ahmed2013peek,bauckhage2013mathematical,matsubara2012rise,yang2010modeling}.

However, the temporal patterns exhibited by cascades over significantly longer time scales is largely unexplored.
Do successful cascades display a long monotonic decline after their
initial peak, or do they exhibit more complex behavior in which
they can {\em recur}, experiencing renewed bursts of popularity
long after their initial introduction?
Anecdotally, many of us have experienced d\'{e}j\`{a} vu when 
a friend shared content we had seen weeks or months ago,
but it is not clear whether these are isolated occurrences
or glimpses into a robust phenomenon.
Resolving these basic distinctions in the long-time-scale
behavior of cascades is crucial to 
understanding the longevity of content beyond its initial popularity, 
and points toward a more holistic view of how content spreads in a network.

\xhdr{The present work: Cascade recurrence}
We perform a year-long large-scale analysis of cascades
of public content on Facebook, measuring them over significantly longer time scales
than previously investigated.
Our first main finding is that recurrence is widespread in 
the temporal dynamics of large cascades.
Among large cascades appearing in 2014, over half come back in one or more subsequent bursts. 
While reshare activity does peak and then drop to very low or even zero levels relatively soon after introduction, the same content can recur after a short or extended lull.

The prevalence of recurrence prompts several 
questions about how and why content recurs. 
Is more broadly or narrowly appealing content more likely to recur?
Does a larger initial burst indicate a greater likelihood of recurrence, 
or does it inhibit subsequent bursts by exposing and thus
satiating many people in the initial wave?
Do different bursts of the same content 
spread in different parts of the network?
Is the second burst a continuation of the initial cascade, or a fresh re-introduction of the content into the network?
Does the media type of the reshared content matter --- 
for example, whether it is a photo or a video?
Finally, how well can one combine such features to predict whether 
a piece of widely reshared
content is likely to experience additional bursts in popularity later on?

We motivate our discussion with an example of content recurrence.
Figure \ref{fig:examplememe} shows an image meme that first became popular 
on Facebook at the end of February 2014, and it 
depicts how the number of reshares of that meme changed over time.
Here, while an initial burst in resharing activity is followed by a gradual decrease, this meme recurred, experiencing multiple resurgences in 
popularity --- first in mid-March, then several times over the next few months.
Perhaps surprisingly, there is little to no resharing between consecutive bursts.
Additionally, multiple near-identical \emph{copies} of this image meme, represented in different colors, are shared in the network.
This distinction between different copies of the same content will prove
important in our later analyses: when a user reshares content through
the reshare mechanism provided by the site, the content continues onward
as the same copy; in contrast, when a user reposts or re-uploads the
same content and thus shares it afresh, this is a new copy.

Figure \ref{fig:communities} sketches the diffusion cascade of this meme, or its propagation over edges in the social network.
As shown in (a), bursts in activity are connected through the same large long-lived cascade and can be traced through the network, from the initial bursts in March (shown in red), to the smaller bursts nearer the end of 2014 (shown in blue).
In (b), where the same network is now colored according to the
copy of the image being reshared, 
different copies of the same content appear at different times, sometimes corresponding to when bursts occur, suggesting that recurrence sometimes occurs from the introduction of new copies.
However, recurrence may also occur as a continuation of a previous copy: 
the copy highlighted in red in (c) experiences an initial burst in March, 
but then resurfaces in popularity later in the year.
Further, we see in (d) that friendship ties exist between even 
the earliest and latest reshares --- the meme appears to be diffusing rapidly, 
but also revisits parts of the network through which it had earlier diffused.

While the meme in our example recurred several times,
are such memes the exception or the norm?
And if such memes are in fact typical, what are the bases
for such robust patterns of recurrence?
To answer these questions, we use a dataset of reshare activity 
of publicly viewable photos and videos on Facebook in 2014.

\xhdr{Characterizing recurrence}
First, we develop a simple definition of a \emph{burst}, 
corresponding informally to 
a spike in the number of reshares over time, that we can use to quantify when recurrence occurs (via multiple observed bursts), and when it does not 
(a single burst).
We show that a significant volume of popularly reshared content recurs (59\% of image memes and 33\% of videos), and that recurring bursts
tend to take place over a month apart from each other.
Recurrence is itself relatively bursty --- rarely do we observe long sustained periods of resharing.

Studying the temporal patterns of recurrence, user characteristics of the resharing population, and the network structure of cascades, we find that 
the recurrence of a piece of content is moderated to a large extent by its 
{\em virality}, or broadness of appeal:
cascades with initial bursts of activity that are larger, last longer, and have a more diverse population of resharers are more likely to recur.
Nonetheless, it is not the cascades that start out the largest or most viral that recur, but those that are moderately appealing.
Specifically, a moderate number of initial reshares, as well as a moderate amount of homophily (or diversity) in the initial resharing population is correlated with higher rates of recurrence.
This lies in contrast to more appealing (or popular) content, where one is likely to see a single large outbreak which results in a large single burst, as well as less appealing content, where one is likely to only see a single small outbreak and thus a smaller single burst.
In the former case, we show evidence that a large initial burst inhibits subsequent recurrence by effectively ``immunizing'' a large proportion of the susceptible population.

While individual copies of content already recur in the network (18\% for image memes and 30\% for videos), the presence of multiple copies catalyzes recurrence, allowing that content to spread rapidly to different parts of the network, significantly boosting the rate of recurrence.
To a smaller extent, the principle of homophily, suggesting that 
people are more likely to share content received from users similar to themselves, also plays a role in recurrence, with user similarity positively correlated with the rate of spreading.

\xhdr{Modeling recurrence}
Motivated by the above picture of recurrence, and inspired by classic epidemiological models of diffusion \cite{newman2002spread} and disease recurrence \cite{altizer2006seasonality,johansen1996simple,olsen1988oscillations}, we present a simple model of cascading behavior that is primarily driven by content virality and the availability of multiple copies, and is able to reproduce the observed recurrence features.
A simulation of this model, which introduces multiple copies of the same content into the network, can cause independent cascades that peak at different times and in aggregate are observed as recurring.
As the virality of the content increases, the shape of a plot of overall reshares in the network over time transforms from a shorter independent single burst, to multiple bursts of differing sizes, to a single large burst of a longer duration.
Replicating our previous findings, increasing virality increases recurrence, up to a point: once a meme has exposed a large part of the network, further recurrence is inhibited.

\xhdr{Predicting recurrence}
Finally, we show how temporal, network, demographic, and multiple-copy features may be used to predict \emph{whether} a cascade will recur, 
if the recurrence will be \emph{smaller} or \emph{larger} than
the original burst, and \emph{when} the recurrence occurs.
We demonstrate strong performance in predicting whether the same content will recur after observing its initial burst of popularity (ROC AUC=0.89 for image memes), as well as in predicting the relative size of 
the resulting burst (0.78).
The time of recurrence, on the other hand, appears to be more unpredictable (0.58).
Features relating to content virality and multiple copies perform best.
Though multiple-copy features account for significant performance in predicting the recurrence of content, we obtain similarly strong performance (0.88) when predicting the recurrence of an individual copy of a piece of content.

Together, these results not only provide the first large-scale study of content recurrence in social media, but also begin to suggest some of the
factors that underpin the process of recurrence.

%% file: 020data.tex

Studying cascade recurrence requires both sufficiently rich data that accurately measures activity throughout a network over long periods of time, as well as a robust definition of what recurrence is.

\subsection{Dataset Description}

In this paper, we use over a year of sharing data from Facebook.
All data was de-identified and analyzed in aggregate.
Facebook presents a particularly rich ecosystem of users and pages (entities that can represent organizations or brands) sharing a large amount of content over long periods of time.

Reliably measuring the spread of content in a network over time is challenging because multiple copies of the same content may exist at any time.
As we will later show, the presence of multiple copies in a cascade is an important catalyst for recurrence.
On Facebook, users and pages may introduce a new copy of the same content by re-posting or re-uploading it; resharing an existing copy instead creates an attribution back to that same copy.
Content may be reintroduced, instead of reshared, for various reasons --- multiple users may have independently discovered the same content, or downloaded and then re-uploaded an image.

To construct a dataset of popularly shared content, we initially selected a seed set of reshared content uploaded to Facebook in March 2014.
We selected the top 200,000 most reshared images, which were publicly viewable, counting only reshares within the 180 days since the image was uploaded,
then used a neural network classifier \cite{krizhevsky2012imagenet} to identify images with overlaid text (i.e., image memes).
One advantage of studying image memes in particular is that the information that these memes transmit is unlikely to change, as opposed to unembellished images which may be used differently (e.g., if the same photo is used to support separate causes).

Next, we tried to identify other copies of content that exist in this seed set.
Beyond exact copies of the same image, many near-identical images, which have slightly different dimensions or introduce compression artifacts or borders, also exist (as seen in Figure \ref{fig:examplememe}).
As such, a binary \textit{k}-means algorithm \cite{gong2011iterative} was used to identify clusters of near-identical images to which each of these candidates belonged, including images beyond the original set.
For each cluster, we then obtained all reshares of images in that cluster that were made in 2014.
To verify the quality of the clustering, we manually examined the top 100 most-reshared copies in each of 100 randomly sampled clusters.
In 94 clusters, all 100 copies were near-identical.
The remaining clusters mainly comprised the same image overlaid with different text.

This sample of resharing activity in 2014 that we use consists of 395,240,736 users and pages that made 5,167,835,292 reshares of 105,198,380 images. These images were aggregated into 76,301 clusters.
Repeating the process above for videos shared on Facebook, we obtain a sample comprising 323,361,625 users and pages that made 2,187,047,135 reshares of 6,748,622 videos, aggregated into 156,145 clusters.
Images, videos, users, and pages that were deleted were excluded from analysis.
On average, each image cluster is made up of 1379 copies of the same content.
Video clusters were smaller, with 43 copies in each cluster on average.

As we only measured reshares for a year, we may only be observing part of a cascade's spread if it began prior to 2014.
Thus, we also considered subsets of each dataset containing only clusters that began in 2014.
We identified these subsets by additionally measuring reshares of content in the three months prior to 2014 (October to December 2013) and excluding clusters where activity was observed during this period.

Though we mainly analyze recurrence at the cluster level, we also investigate the recurrence of individual copies by studying the top 100,000 individually most reshared copies in each dataset.

\begin{figure}[tb]
    \centering
    \includegraphics[width=0.5\textwidth]{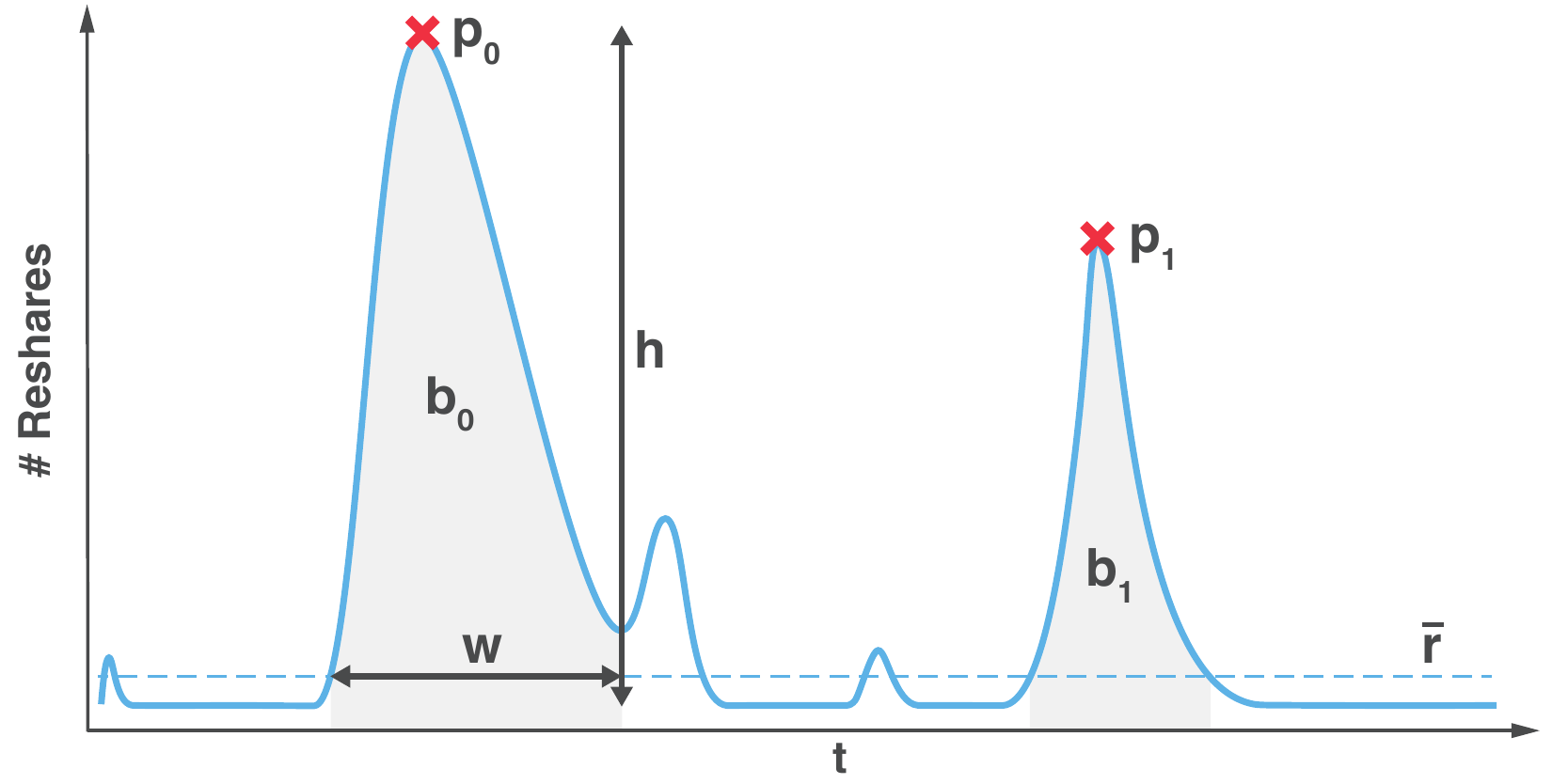}
    \caption{Recurrence occurs when we observe multiple peaks ($p_0$, $p_1$, red crosses) in the number of reshares over time. Bursts ($b_0$, $b_1$) capture the activity around each peak.}
    \label{fig:definition}
\end{figure}

\begin{figure}[tb]
    \centering
    \includegraphics[width=0.5\textwidth]{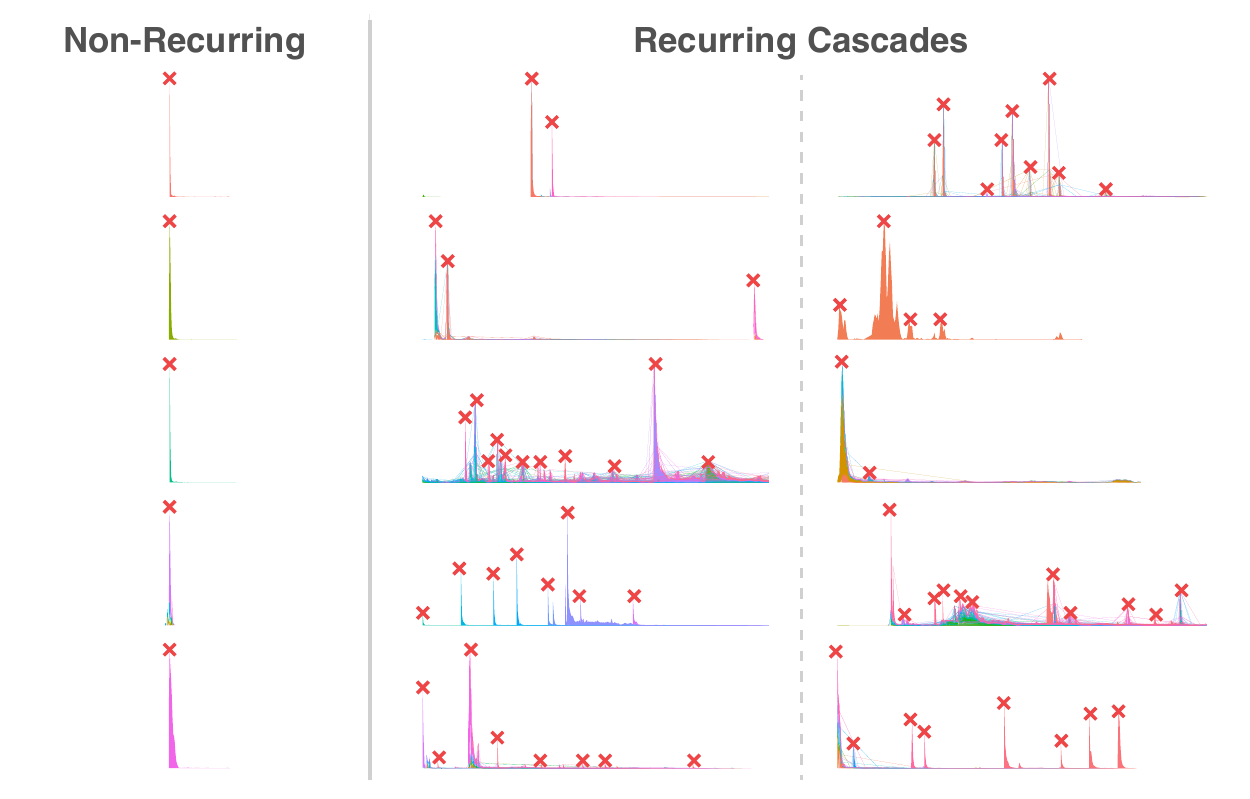}
    \caption{Examples of time series of recurring and non-recurring cascades over a year, colored by copy. Identified peaks are marked with red crosses; the number of reshares is normalized per cascade.}
    \label{fig:timeseries}
\end{figure}

\begin{table*}[tb]
\small
\centering
\ra{1.3}
\begin{tabular*}{\textwidth}{@{\extracolsep{\fill}}lllllll}\toprule
   & \textbf{\# Clusters} & \textbf{Copies/Cluster} & \textbf{Prop. Recurrence} & \textbf{\# Peaks} & \textbf{Days Observed} & \textbf{Days betw. 1st/2nd Burst} \\
  \hline
  Image Memes & 51,415 \deem{(76,793)} & 523 \deem{(1378)} & 0.40 \deem{(0.59)} & 2.3 \deem{(4.6)} & 202 \deem{(280)} & 31 \deem{(32)} \\
  \hline
  Videos & 149,253 \deem{(156,145)} & 13 \deem{(43)} & 0.30 \deem{(0.33)} & 1.6 \deem{(2.0)} & 170 \deem{(182)} & 47 \deem{(44)} \\
\bottomrule
\end{tabular*}
\caption{Recurrence occurs in a large proportion of popular image memes and videos shared on Facebook. We note in parentheses statistics computed on all cascades, as opposed to the cascades that began in 2014 whose initial spread we can observe.}
\label{tab:recurrence}
\end{table*}

\subsection{Defining Recurrence}
\label{sec:definition}
In this work, we define recurrence relative to \emph{peaks} and \emph{bursts} in popularity over time.
In practice, almost all popular content on Facebook experiences at least one peak in popularity.
If content peaks in popularity more than once, we say that it \emph{recurs}.

To identify these peaks, and thus whether a cascade recurs, we measure the number of reshares of content over time.
Figure \ref{fig:timeseries} shows several examples of recurring memes.
Empirically, reshare activity is varied across different content but is generally bursty, with long periods of inactivity between peaks.
As recurrence occurs over a long amount of time, we discretize time into days.

Intuitively, a recurrence occurs when a \emph{peak} is observed in the time series. Not only should these peaks be relative outliers on a timeline, but they should also last for a significant amount of time. Further, we should be able to tell these peaks apart from each other.
Motivated by this intuition, suppose we observe a meme for $t$ days. Let $r_i,~i \in \{1,2,...,t\}$ be the number of reshares observed on day $i$.
We parameterize recurrence using four variables --- $h_0$, $m$, and $w$ place constraints on identified peaks, and $v$ places a constraint on the ``valley'' between peaks.
Specifically, the height $h$ of each peak must be at least $h_0$ and at least $m$ times the mean reshares per day $\bar r$ (Figure \ref{fig:definition}).
Additionally, a peak day must be a local maximum within $\pm w$ days. 
Finally, between any two adjacent peaks $p_i$ and $p_{i+1}$, the number of reshares must drop below $v \cdot \min \{r_{p_i}, r_{p_{i+1}}\}$.
We call the area around the peak a \emph{burst} ($b_0$, $b_1$ respectively for $p_0$, $p_1$ in Figure \ref{fig:definition}), whose duration or width $w$, is defined as the sum of the number of days the number of reshares is increasing before $p_i$ and falling after $p_i$, while remaining above $\bar r$.
There is a one-to-one correspondence between peaks and bursts.

In practice, we set $h_0$=10, $m$=2, $w$=7, and $v$=0.5 so that each burst is relatively well-defined.
The red crosses in Figure \ref{fig:timeseries} show the identified peaks under this regime.
While this definition does not strictly minimize activity between bursts, empirically, activity does drop significantly (and in many cases, falls to zero) in between bursts.
Stricter definitions that reduce the number of identified peaks (e.g., requiring a well-defined ``valley floor'' between two peaks, or increasing $h_0$ or $m$) also resulted in qualitatively similar findings.
The approach we take is fairly rudimentary; future work may involve developing more specific definitions of recurrence which take into account the shape of resulting bursts.

%% file: 030characteristics1.tex

We first introduce recurrence at a high level, showing that it is both common and bursty, with the same content sometimes resurfacing multiple times.
We then discuss four important classes of observations that we later draw on to model and predict recurrence:
\begin{itemize}[noitemsep]
    \item Temporal patterns: cascades with longer initial bursts, but a moderate number of reshares, are more likely to recur.
    \item Sharer characteristics: recurring and non-recurring cascades differ in demographic makeup, and moderate diversity in the initial sharing population encourages recurrence. Further, changes in homophily in the network affect the speed at which content spreads, and hence burstiness.
    \item Network structure: bursts in a cascade occur in different, but nonetheless connected parts of the network. Also, large initial bursts tend to exhaust the supply of susceptible users, potentially accounting for why moderate, but not high cascade volume or diversity results in greater recurrence.
    \item Catalysts of recurrence: the availability of multiple copies in the network may catalyze recurrence. Still, neither does the presence of multiple copies suggest that recurrence is entirely an externally-driven phenomenon, nor is it a necessary condition for recurrence.
\end{itemize}

In the remainder of this paper, we report results primarily on image memes, and note any salient differences with videos.
All differences reported are significant at \pvallow~using a \textit{t}-test unless otherwise noted.

\subsection{Recurrence is common}
Once introduced on Facebook, popular content continues spreading for a long time.
On average, the maximum time between reshares of the same content is 280 days.
But rather than being shared at a constant rate (among popularly reshared content, less than 1\% of memes have no discernible peak), resharing tends to be bursty, with bursts typically separated by substantial periods of relative inactivity.
A mean of 32 days separates the initial and subsequent bursts for image memes (Figure \ref{fig:days_between_dist}).

Previously, we defined recurrence as observing multiple peaks in the number of reshares observed over time, 
and non-recurrence as observing only a single peak.
Over these long periods of time, 59\% of popular image memes recur.
In fact, a significant proportion of these cascades experience resurgences in popularity (Figure \ref{fig:peak_dist}), and may even have experienced bursts prior to our observation window.
If we limit the sample to the set of image memes which \emph{began} spreading in 2014, 40\% of these memes recur (Table \ref{tab:recurrence}).

%% file: 035characteristics2.tex

\subsection{Temporal patterns}
Cascades with larger initial bursts of activity that last longer are more likely to recur, suggesting that more viral, or appealing cascades are more likely to recur.
However, it is not the most popular cascades that recur the most, but those that are only moderately popular --- while recurrence initially increases with the size of the initial peak, it subsequently decreases.

\begin{figure*}[tb]
        \centering
        \begin{subfigure}[b]{0.32\textwidth}
                \includegraphics[width=\textwidth]{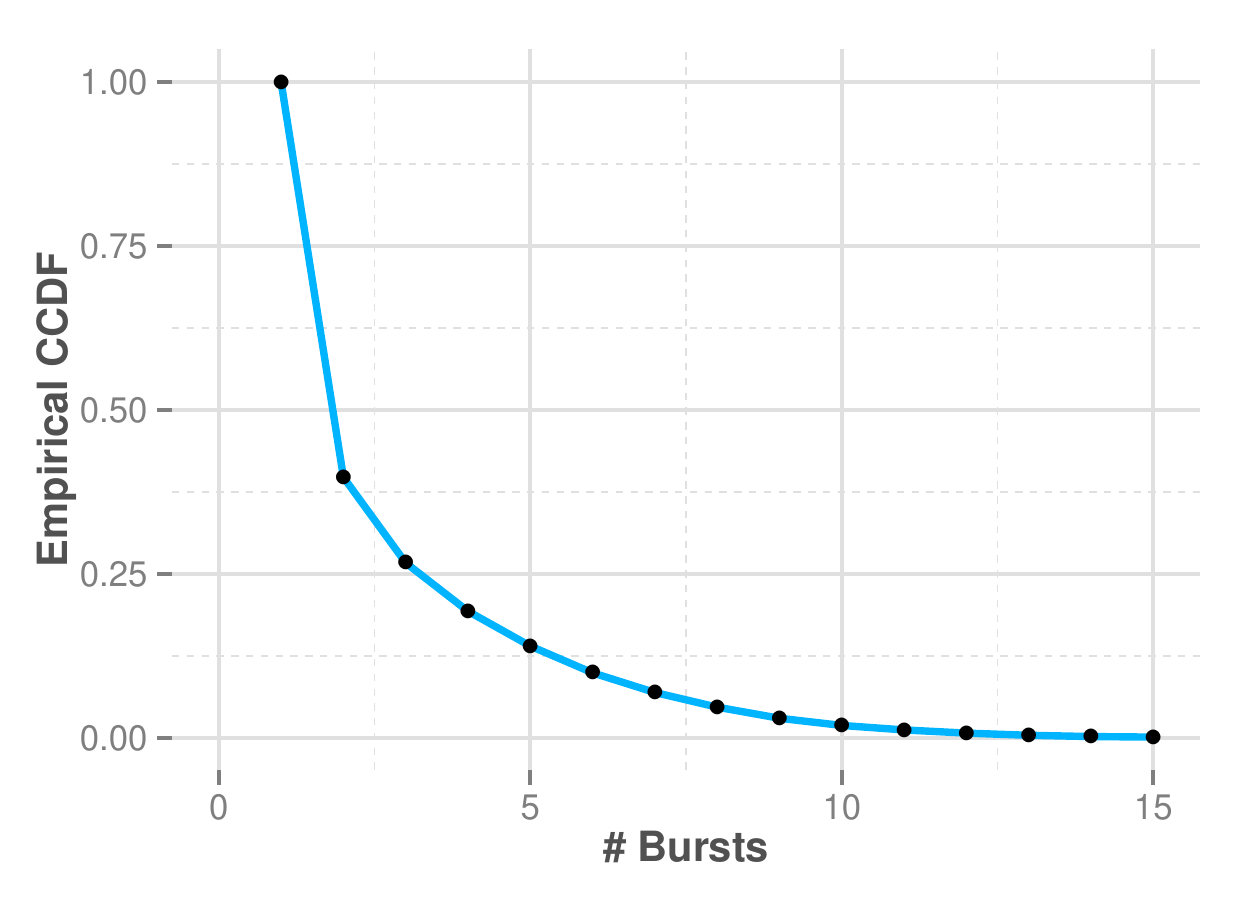}
                \caption{Number of Peaks}
                \label{fig:peak_dist}
        \end{subfigure}
        \begin{subfigure}[b]{0.32\textwidth}
                \includegraphics[width=\textwidth]{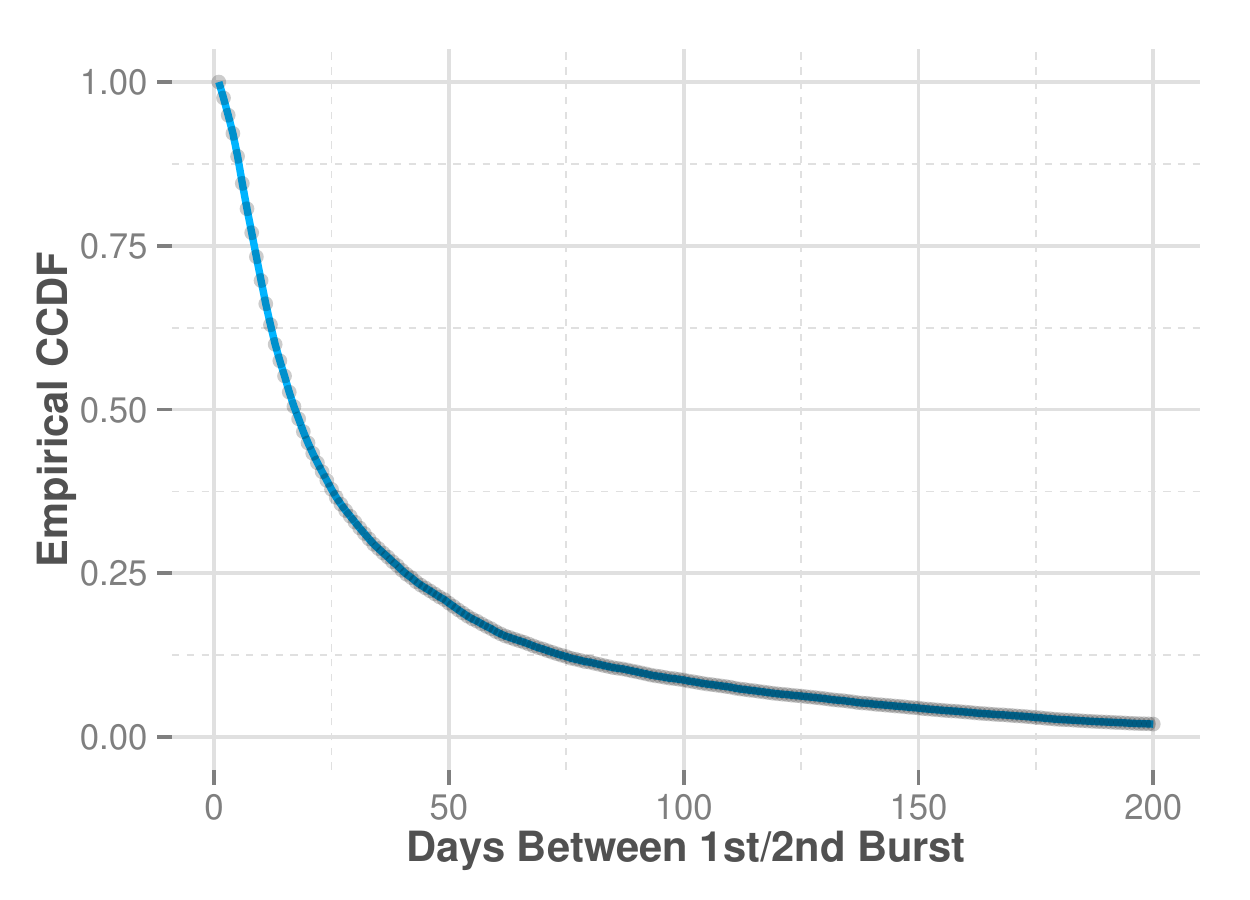}
                \caption{Days Between Bursts}
                \label{fig:days_between_dist}
        \end{subfigure}
        \begin{subfigure}[b]{0.32\textwidth}
                \includegraphics[width=\textwidth]{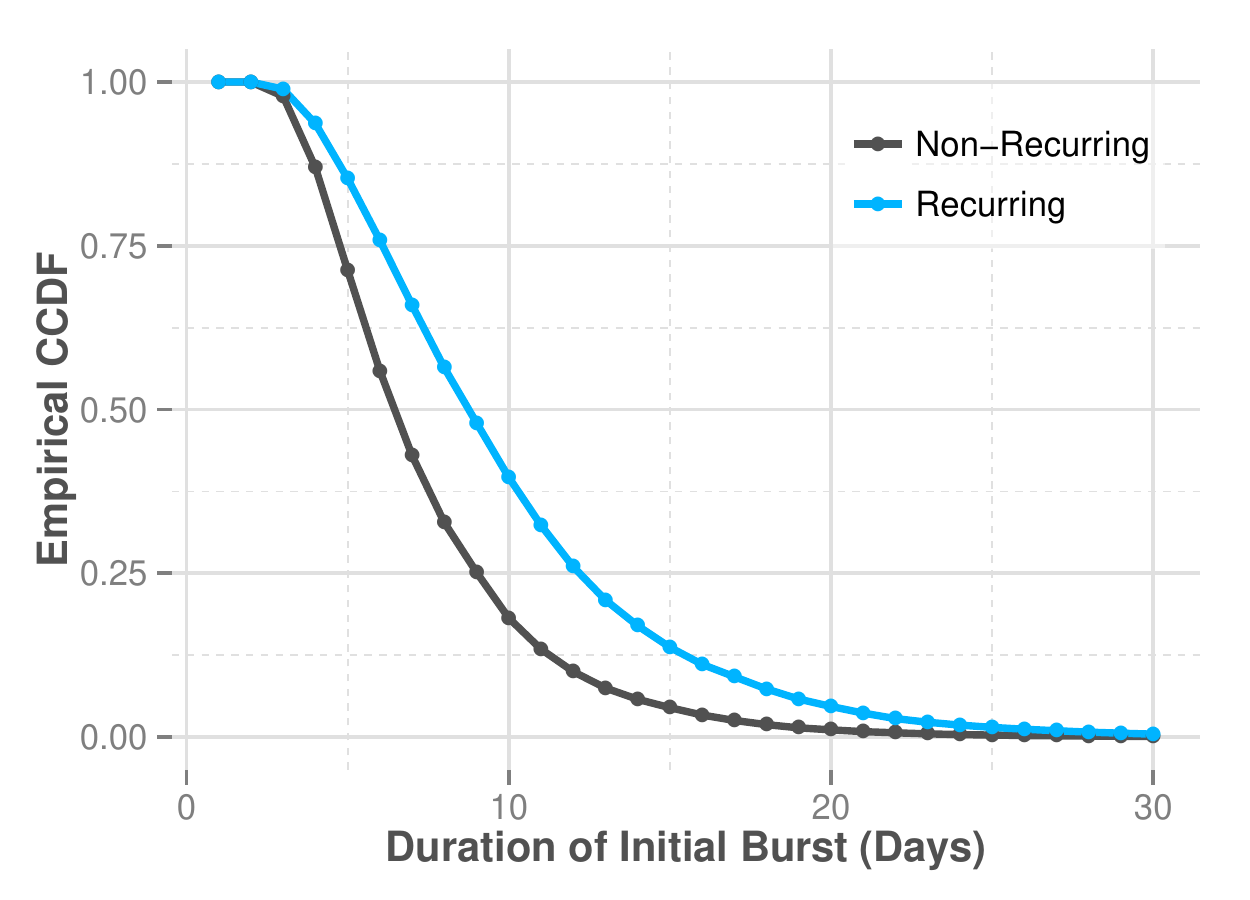}
                \caption{Cascade Duration}
                \label{fig:days_0_dist}
        \end{subfigure}
        \caption{(a) 40\% of cascades that began in 2014 came back, and (b) over 30\% of recurring cascades only resurfaced after a month or more. (c) Further, the initial burst of a recurring cascade tends to last longer than that of a non-recurring cascade.}\label{fig:peaks}
\end{figure*}

\xhdr{Recurring cascades have larger, longer-lived initial bursts}
The initial burst of a cascade is already indicative of recurrence.
Recurring cascades start out larger (15,547) and initially last longer (9.3 days) than non-recurring cascades (6128 reshares, 6.9 days), spending more time ``building up''  (Figure \ref{fig:days_0_dist}) and ``winding down''.
The greater initial popularity of recurring cascades suggests that more viral cascades are more likely to recur, but is this the case?

\begin{figure*}[ht]
        \centering
        \begin{subfigure}[b]{0.32\textwidth}
                \includegraphics[width=\textwidth]{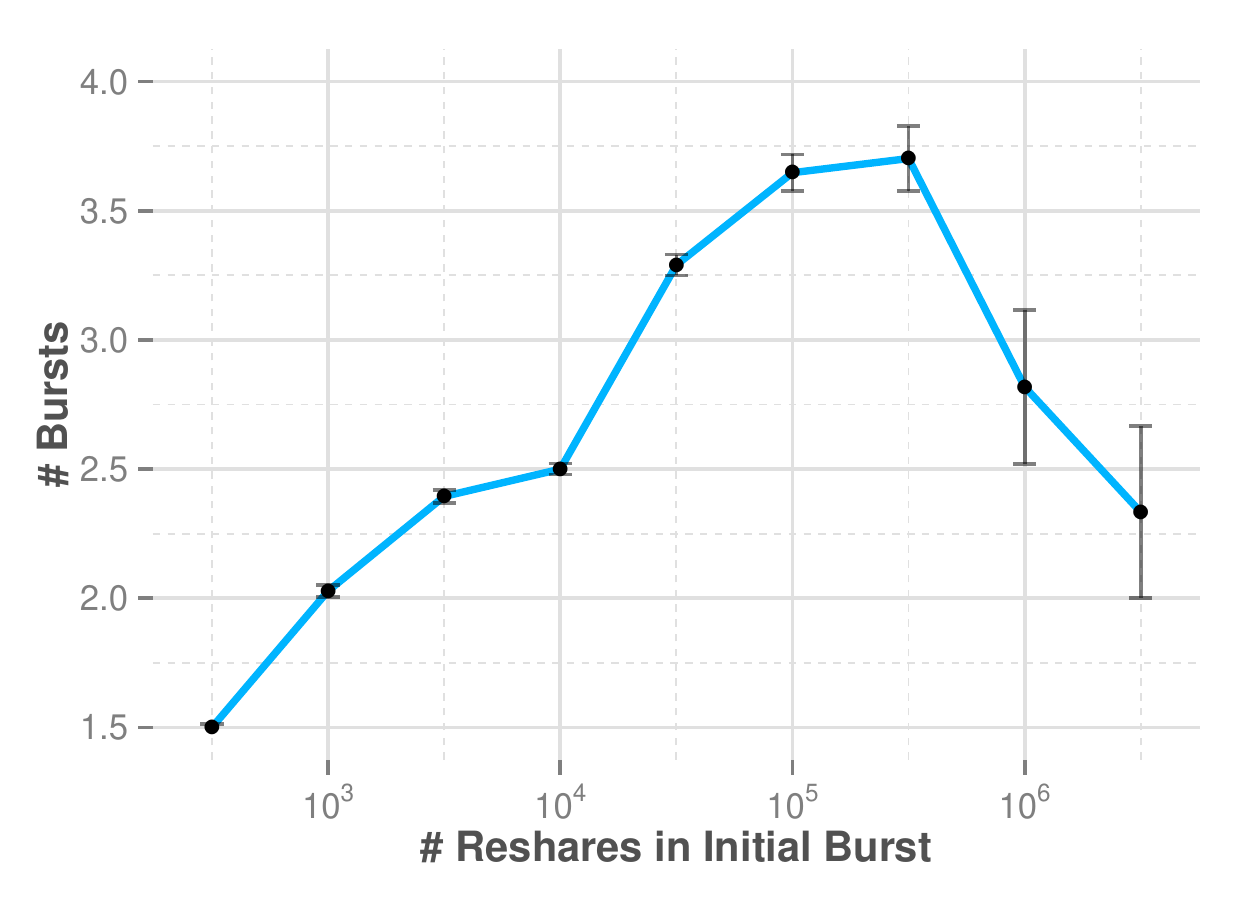}
                \caption{\# Bursts vs. \# Reshares}
                \label{fig:imax_peaks_reshares}
        \end{subfigure}
        \begin{subfigure}[b]{0.32\textwidth}
                \includegraphics[width=\textwidth]{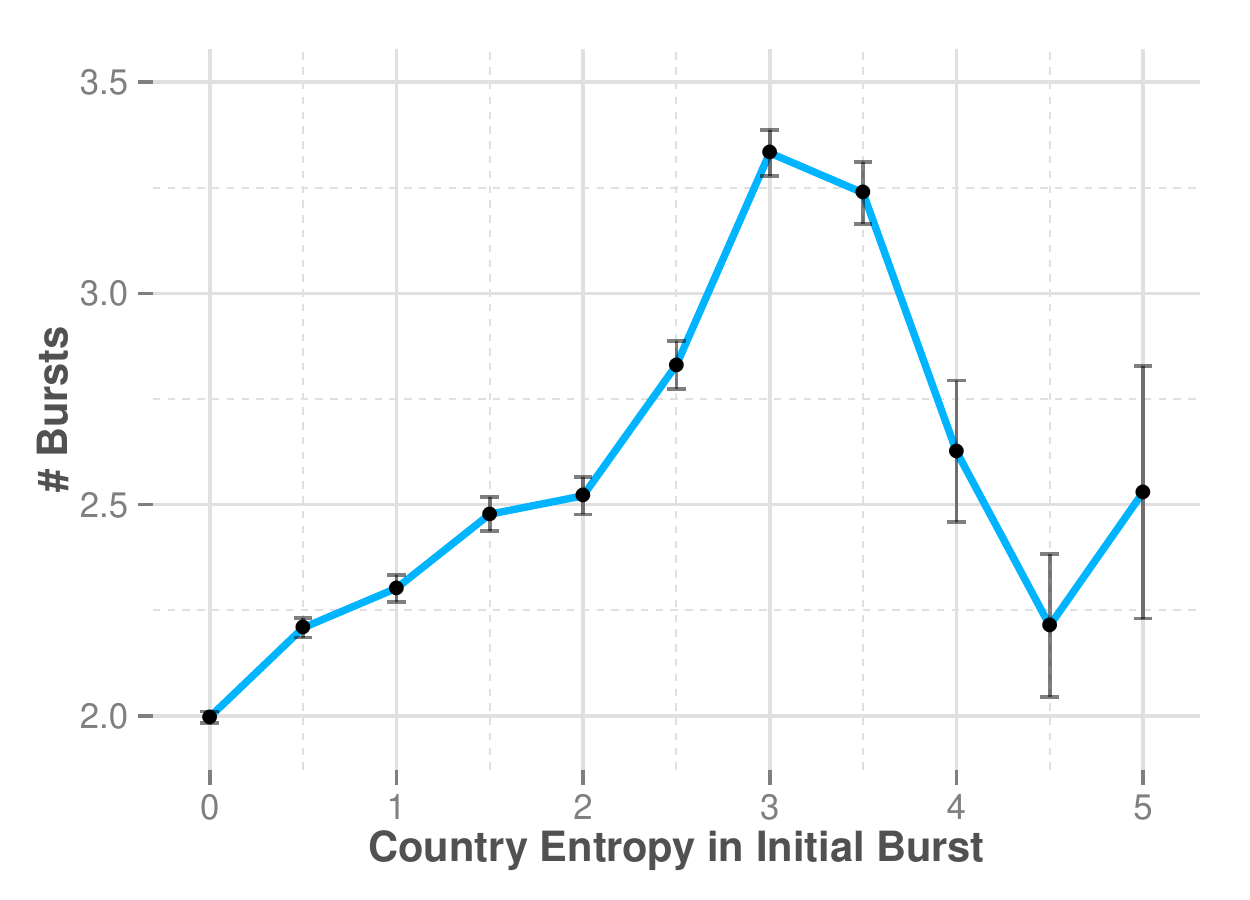}
                \caption{\# Bursts vs. Country-Entropy}
                \label{fig:demographics_country_entropy}
        \end{subfigure}
        \begin{subfigure}[b]{0.32\textwidth}
                \includegraphics[width=\textwidth]{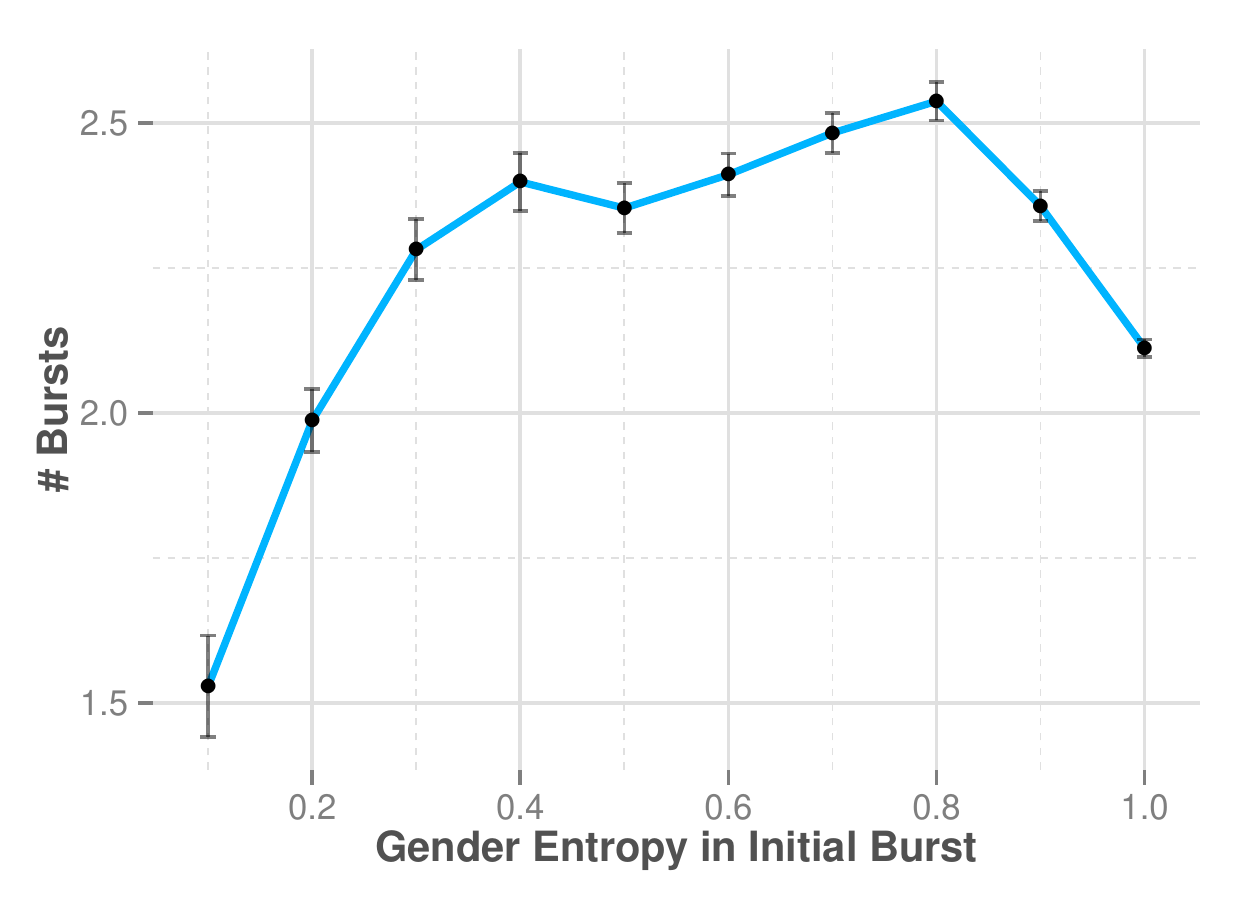}
                \caption{\# Bursts vs. Gender-Entropy}
                \label{fig:demographics_gender_entropy}
        \end{subfigure}
        \caption{(a) A moderate number of reshares results in more recurrence. (b), (c) Similarly, recurrence is more likely when the entropy of the distribution of users across countries, as well as gender, is moderate.}\label{fig:demographics}
\end{figure*}

\xhdr{Recurring content is moderately popular}
Plotting the total number of reshares in the initial burst against the subsequent number of bursts observed, rather than the number of reshares monotonically increasing or decreasing the rate of recurrence, we observe a striking interior maximum at approximately 10\textsuperscript{5} reshares for both image memes and videos (Figure \ref{fig:imax_peaks_reshares}).
Neither the initially best-performing (or most viral), nor poorest-performing (or least viral) cascades tend to resurface.
In the former case, a single large burst tends to dominate with smaller bursts after; in the latter case, a small number of small bursts is typically observed.

\begin{figure}[tb]
        \begin{subfigure}[b]{0.49\columnwidth}
                \includegraphics[width=\textwidth]{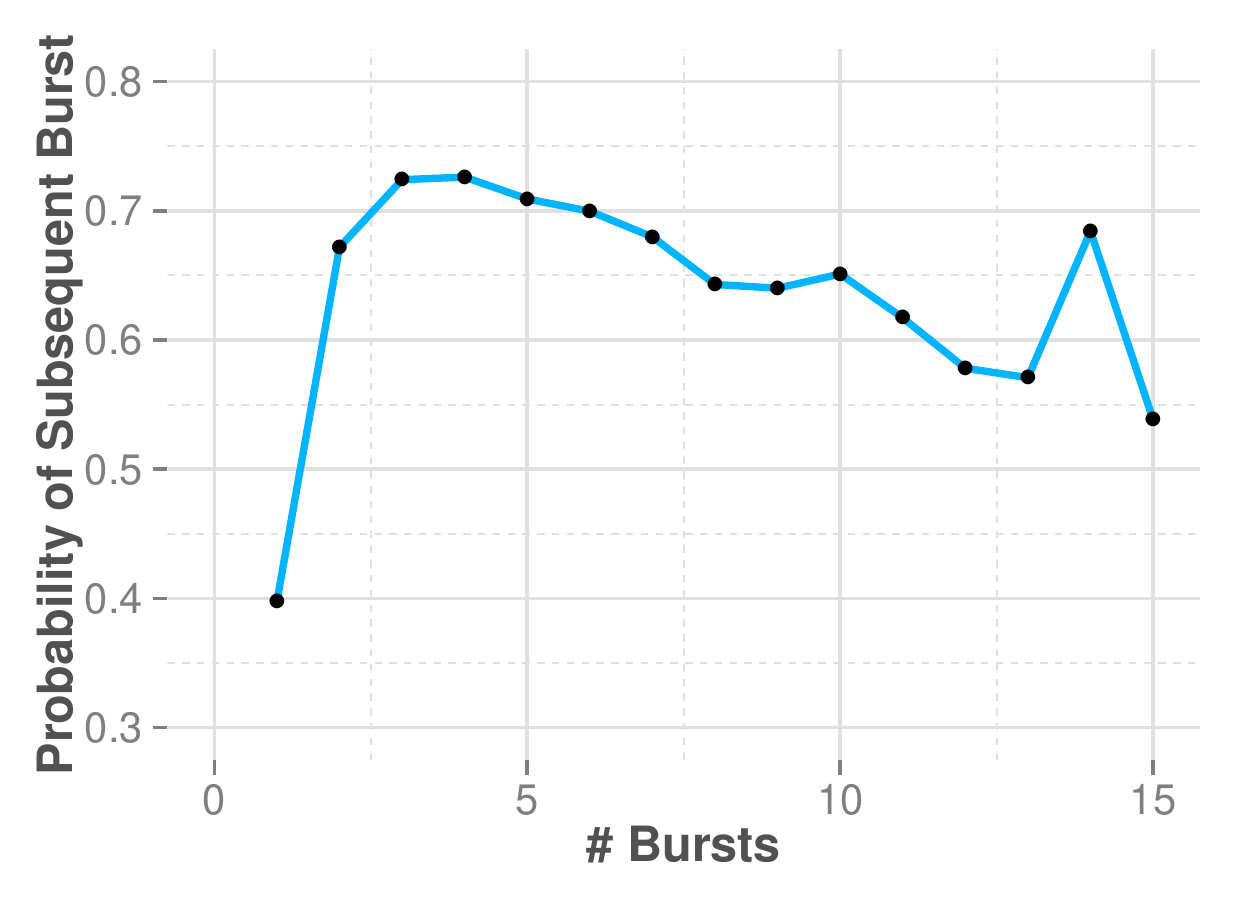}
                \caption{Probability of Subsequent Bursts}
                \label{fig:pr_recurrence}
        \end{subfigure}
        \begin{subfigure}[b]{0.49\columnwidth}
                \includegraphics[width=\textwidth]{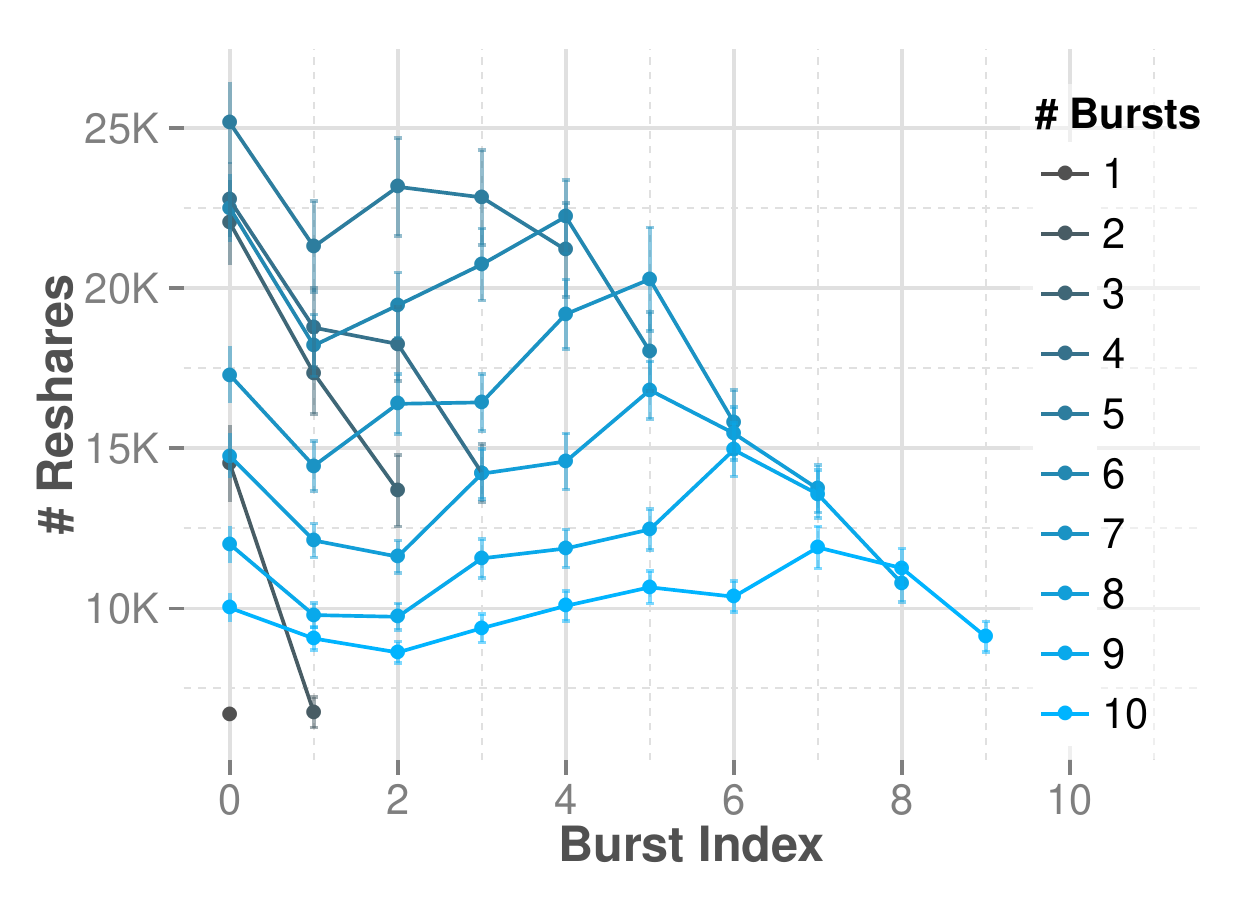}
                \caption{\# Reshares in Each Burst}
                \label{fig:peak_index_reshares_separated}
        \end{subfigure}
        \caption{(a) The probability of subsequent recurrences increases after the initial recurrence. (b) Cascades that recur less tend to have bursts that diminish in size over time, while those that recur more tend to have a stable burst size.}\label{fig:peaks2}
\end{figure}

\xhdrr{They keep coming back!}
While most of our analyses focus on the initial burst and subsequent recurrence, several general trends arise as more recurrence is observed:
\begin{itemize}[noitemsep]
    \item Once a cascade has recurred, it is more likely to resurface again.
    The probability of recurrence jumps from 0.40 initially, to 0.60 for subsequent recurrences before gradually decreasing (Figure \ref{fig:pr_recurrence}).
    This observation parallels prior work showing that the prior popularity of Youtube videos predicts their future popularity \cite{borghol2012untold}.
    In fact, for 26\% of all image meme cascades, we observe resharing activity on the first and last day of our observation period.
    These image memes may be ``evergreen'', tending to continuously recur.
    \item For cascades that recur less, subsequent bursts tend to be smaller; for cascades that recur more, subsequent bursts are more similar in size (Figure \ref{fig:peak_index_reshares_separated}), suggesting that they depend less on external factors (e.g., breaking news) to spread.
    \item Subsequent recurrences are briefer than their predecessors. Burst duration monotonically decreases from a mean of 7.6 days for the first burst to 6.3 for the tenth.
    \item On average, the lull between recurrences is substantial, with bursts happening an 28 to 32 days apart for image memes, and 30 to 44 days apart for videos.
    Again, these long periods between bursts suggest that recurrence can only be observed over substantial periods of time.
\end{itemize}

\subsection{Sharer characteristics}
People who participate in recurring cascades differ significantly from those who participate in non-recurring cascades.
While a diverse user population encourages recurrence, moderately diverse cascades recur the most.
Homophily, the concept that similar people are likely to share the same content, also affects how quickly content spreads, suggesting that it modulates recurrence.

\xhdr{Demographics vary with recurrence}
For recurring cascades, the average age of people participating in the initial burst is lower (40 vs. 42), but the proportion of women is higher (65\% vs. 58\%). The latter observation corroborates previous work that showed a correlation with eventual cascade size \cite{cheng2014can}.

Demographics also change across bursts.
In the case of image meme cascades, the mean age changes by 2.7 years, and the proportion of women by 6.1 percentage points (in absolute terms).
The same content may become popular in different parts of the world at different times, resulting in recurrence: 13\% of the time, the majority of people in the initial two bursts come from different countries.

\xhdr{Diversity encourages recurrence}
We now turn our attention to the diversity (or homophily) of people who take part in a cascade.
We quantify homophily in the network by measuring the entropy of the distribution of demographic characteristics.
A low entropy in the distribution of countries users are from (or country-entropy) corresponds to high homophily, suggesting that a majority of sharers belong to a small number of countries. On the other hand, a high country-entropy suggests that the countries sharers belong to are more diverse and distributed more evenly.

It is not \textit{a priori} clear whether homophily encourages or inhibits recurrence.
Homophily within a community, meaning that connected users are receptive to sharing the same content, may help a cascade gain the initial traction it needs to spread, but may also result in the content getting ``trapped'' in a local part of the network.
In contrast, diversity in the users sharing that content suggests it has wider appeal and might come back, but may also result in only a single burst if the initial spread overwhelms the network.

We find that diversity in the country distribution is predictive of recurrence.
Controlling for the duration ($w$), peak height ($h$), and the number of reshares in the initial bursts of recurring and non-recurring cascades \cite{rosenbaum1983central}, a Wilcoxon Signed-rank test shows that a higher country-entropy is indicative of recurrence (\textit{W}>10\textsuperscript{8}, \textit{p}\textless10\textsuperscript{-10}, effect size \textit{r}=0.19).
Thus, if the initial burst of a cascade occurs in more countries, it is more likely to recur.
Higher gender-entropy (i.e., greater gender-balance) also predicts recurrence, but its effect is weaker (\textit{W}>10\textsuperscript{8}, \textit{p}\textless10\textsuperscript{-2}, \textit{r}=0.02).
The effect of age is inconsistent across image memes and videos.

\xhdr{Recurring content is moderately diverse}
Again, it is not the most diverse populations that bring about recurrence: a moderate country-entropy of approximately 3.0 in the initial burst of a cascade results in the most recurrence (Figure \ref{fig:demographics_country_entropy}).
An interior maximum can also be observed with respect to the gender-entropy of the initial burst (Figure \ref{fig:demographics_gender_entropy}).
These results, combined with the previous observation of a similar interior maximum with respect to the initial number of reshares, suggests that the virality of content plays a significant role in recurrence.

\xhdr{Cascades spread quickly in pockets of homophily}
The virality of a cascade and homophily in the network are closely related, and perhaps represent two perspectives on the spread of content.
Greater virality enables content to appeal to a larger population; more homophily suggests that  receptive users are closer in the network.
In fact, homophily in the network modulates the speed of resharing (and thus bursts in a cascade).
If we measure the average country-entropy of a sliding window of 100 reshares ordered in time and the time elapsed, and then compute the average correlation, we find a slight positive correlation between the two (0.08), suggesting that homophily among those sharing results in faster resharing, and hence burstiness.
Gender-entropy and age-entropy are also positively, but more weakly correlated with burstiness (0.06 and 0.04 respectively).
One potential cause of this is that pages, with their substantial and homophilous followings, are driving resharing. The more pages that share content, the more homophilous the set of potentially reachable people, and thus the quicker content is reshared.
However, as the proportion of reshares attributable to pages increases, the entropy in these demographic characteristics instead increases, an effect opposite to what we observe.

\subsection{Network structure}
The initial bursts of recurring cascades tend to be better connected.
Further, successful recurrence tends to occur in different, but not disconnected parts of the network.
Considering the people potentially exposed in each burst beyond the resharers, large initial cascades may exhaust the population of susceptible people in the network, a fact that will subsequently become important in explaining the mechanism of recurrence.

\xhdr{Bursts of recurring cascades are internally more connected}
More people and pages share in the initial burst of recurring cascades than non-recurring cascades (15,050 users and 59 pages, and 5855 users and 24 pages respectively).
To measure connectivity within a burst, we used the induced subgraph $G_0$ of the Facebook network made up of the people and pages resharing in the initial burst. The subgraph includes two kinds of edges: friend edges between people, and follow edges between people and the pages they like.
On this subgraph, people in the initial burst of recurring cascades have an average of 3.4 connections to other people and pages in the same burst, relative to 3.1 connections in non-recurring cascades, suggesting that the initial bursts of recurring cascades are slightly more connected.

\xhdr{Subsequent bursts happen in different parts of the network}
Bursts of a cascade are separate in time, but may overlap in terms of sharers, or be connected via friend and follow edges.

To start, the sharer overlap between bursts is small.
For recurring cascades, an average of 15,050 people and 59 pages make up the initial burst, and 8892 people and 28 pages the subsequent burst.
Comparing people and pages across these bursts, Jaccard similarities of 0.02 (i.e., 2\% of people share the same content in both bursts) and 0.03 mean that bursts have very little direct overlap.

We also find evidence of community structure within bursts by considering whether the second burst is proximate in the network to the first.
Even if individuals in the second burst do not repost content a second time, they may not be far removed in the social network from someone in the first burst.
Here, we instead consider the induced subgraph $G_{0+1}$ of the individuals and pages who reshared in the initial two bursts for each cascade.
If these bursts correspond to communities within the network, we would expect more edges within bursts than between them.
An average of 17,457 friend and 17,242 follower edges exist in the first burst; 10,094 friend and 6,310 follower edges exist in the second.
Note that these communities are very sparse.
A person within a burst has a connection to an average of 3.2 friends or pages within the same burst,
indicating that memes tend to diffuse out through the network rather than stay within a narrow community.
Still, across bursts, we observe an average of 8,273 friend and 4,755 follower edges, resulting in an average of 1.4 connections to friends and pages in a different burst, indicating that these bursts are somewhat separated.

\xhdr{Large initial bursts exhaust the supply of susceptible people}
As noted above, the second burst in a cascade has fewer sharers.
Intuitively, people may tire of content they have seen before, but is this the case?
Studying overlap a third way, we look at the susceptible populations of the initial and subsequent bursts of a cascade.
We approximate these populations by considering people who could have been exposed through their connections (friend and follow edges) to those who shared content in these two bursts.

On average, 6.4 million unique individuals are potentially exposed in the initial burst, with recurring cascades, having a greater number of initial resharers, having greater reach (8.2 million vs. 4.0 million for non-recurring cascades). For recurring cascades, the potential reach of the second burst is smaller, but still sizable at 6.8 million. Comparing the sets of individuals exposed in the first two bursts of recurring cascades, we obtain a Jaccard similarity of 0.15, indicating this second burst is mostly reaching a different set of people, but where some people will have seen the same content twice, creating a sense of d\'{e}j\`{a} vu for many.

In particular, 28\% of people who are potentially exposed in the second peak would have also been exposed in the first. 
This high overlap could be a contributing factor to the second peak being smaller.
Further, the proportion of exposed individuals in the second peak is positively correlated ($\rho$=$0.39$) with the size of the first peak, meaning that the larger the initial peak, the more likely that those exposed in the second peak have previously seen the content.
In other words, in the case of large initial bursts, subsequent bursts are likely reaching a similar part of the network.

\subsection{Catalyzing recurrence}
As shown in Figures \ref{fig:examplememe} and \ref{fig:communities}, cascades are made up of reshares of multiple copies of the same content, and the presence of these copies can help catalyze recurrence.
Still, neither are copies the only cause of recurrence (recurrence is substantial even with a single copy), nor must they be independently or externally introduced (many later copies are attributable to previously seen copies).

\xhdr{Cascades whose reshares are divided across multiple copies tend to recur}
Recurring cascades are made up of more copies than non-recurring cascades (2277 vs. 93).
Reshares are also more spread out across multiple copies in the former case (841 and 3445 reshares per copy for recurring and non-recurring cascades respectively), suggesting recurrence may be characterized by multiple smaller outbreaks.
The most reshared copy accounts for 72\% of reshares in the initial burst for recurring cascades, and 93\% for non-recurring cascades.
Altogether, the substantial differences here suggest the strong predictive power of these characteristics.

\xhdr{The appearance of new copies correlates with recurrence}
Further, the introduction of new copies and the number of reshares over time is significantly correlated (Pearson's \textit{r}=0.66), suggesting that the appearance of new copies causes bursts, and thus recurrence.
On a related note, prior work showed that reposting content helps make it popular \cite{stoddard2015popularity}.

\xhdr{Copies are not the only cause of recurrence}
Nonetheless, not all copies burst (only 6\% are reshared at least 10 times on any single day), and not all bursts are caused by new copies, as we will later show.
And while correlations between the number of copies and other characteristics such as duration and country-entropy also exist, when we control for the number of copies in the initial bursts of recurring and non-recurring cascades \cite{rosenbaum1983central}, all previously observed differences in the temporal, sharer, and network characteristics of these cascades still hold (\textit{W}\textgreater10\textsuperscript{8}, \pvallow, mean effect size \textit{r}=0.08).
Comparing recurring and non-recurring cascades with similar numbers of copies in their initial bursts, the initial bursts of recurring cascades are still larger, longer-lived, and more diverse.
In all, this suggests that recurrence is not simply caused by distinct copies of the same content spreading through the network, but is a result of a more complex phenomenon which we explain in Section~\ref{sec:model}.

\xhdr{A majority of copies are internal to the network}
Still, where do these copies come from, and are they internal or external to the network?
By using the network to identify friends and pages who may have previously shared a different copy of some content, we can attribute 75\% of newly uploaded copies to previously seen copies in the network (this approach roughly estimates content-copying that occurs within Facebook, as users who share a new copy may not have seen a friend's shared copy).
This suggests a nuanced approach to studying recurrence --- external sources may drive some of the introduction of new copies to a social system, but a large proportion of activity, which we can study, occurs within the network.

\xhdr{Pages may also catalyze recurrence}
Pages are responsible for a large proportion of highly-reshared copies (over 70\% of reshares are attributable to page-created copies in the second burst of recurring cascades).
In recurring cascades, pages tend to re-upload, rather than reshare content, doing so 50\% of the time, as opposed to 2\% for users.
Further, the most popular copy in the second burst is likely to have been created by a page (70\%).
Given the relatively higher degree of pages, which tend to have tens of thousands of followers, as opposed to users who typically only have hundreds of friends, pages may spark recurrence by posting a new copy of the same content, rapidly exposing a number of followers to it.

\xhdrr{Individual copies recur too!}
Recurrence of the individually most popular copies in our datasets, while lower than when copies are studied in clusters, is still substantial (18\%).
These individual copies last a significant amount of time (261 days), with bursts further apart (41 days).
Like cascades of multiple copies, the initial bursts of recurring individual-copy cascades are larger and longer-lived than those of non-recurring cascades, with later bursts occurring in different parts of the network.
Recurrence of the same copy can also be observed within clusters --- 22\% of the time, the most reshared copy in a burst was also most reshared in a previous burst.

%% file: 040model.tex

Tying our observations together, we present an overall picture of the mechanisms of recurrence, then suggest a model of recurrence which we evaluate through simulations on a real social network.

\subsection{Why do cascades recur?}

Our findings as a whole suggest a model of recurrence where virality is a primary factor, and where the availability of multiple copies can help spark recurrence.

\xhdr{Virality plays a primary role in recurrence}
Virality, or broadness of appeal, affects recurrence: cascades with initial bursts that are larger, last longer, and are more demographically diverse are more likely to recur.
Specifically, \emph{moderately} popular and diverse cascades are \emph{most} likely to recur.
While recurrence typically occurs in different parts of the network, the larger the initial burst of a cascade, the larger the proportion of the potentially exposed population in the subsequent burst that was already previously exposed.
This observation, coupled with the fact that users tend not to reshare the same content multiple times, suggests that large initial bursts inactivate a significant portion of the network, inhibiting a cascade's future spread.
Our subsequent simulations show more clearly that this may indeed happen as the initial burst grows large.

\xhdr{Multiple copies in the network help spark recurrence}
Bursts in a cascade are separated by relatively long periods of inactivity.
By studying the availability of multiple copies of the same content, we find that these copies can act as catalysts for recurrence in different parts of the network.
Indeed, multiple introductions of the same content correlate with recurrence.
However, while more copies initially increases the chance of recurrence, they are not the only cause of it; recurring and non-recurring cascades with similar numbers of copies differ significantly in virality.
Moreover, multiple copies do not explain the substantial recurrence of individual copies.
To a lesser extent, we also discover that homophily in the network affects the speed of the spread of a cascade in a network.

Together, moderate content virality and the presence of multiple copies results in recurrence.
While the likelihood of recurrence does increase with the number of copies (or potential ``sparks''), we can still observe an interior maximum in how recurrence varies with the number of reshares after fixing the number of copies, where a moderate number of reshares results in the most recurrence.

\begin{figure}[tb]
    \centering
    \includegraphics[width=0.5\textwidth]{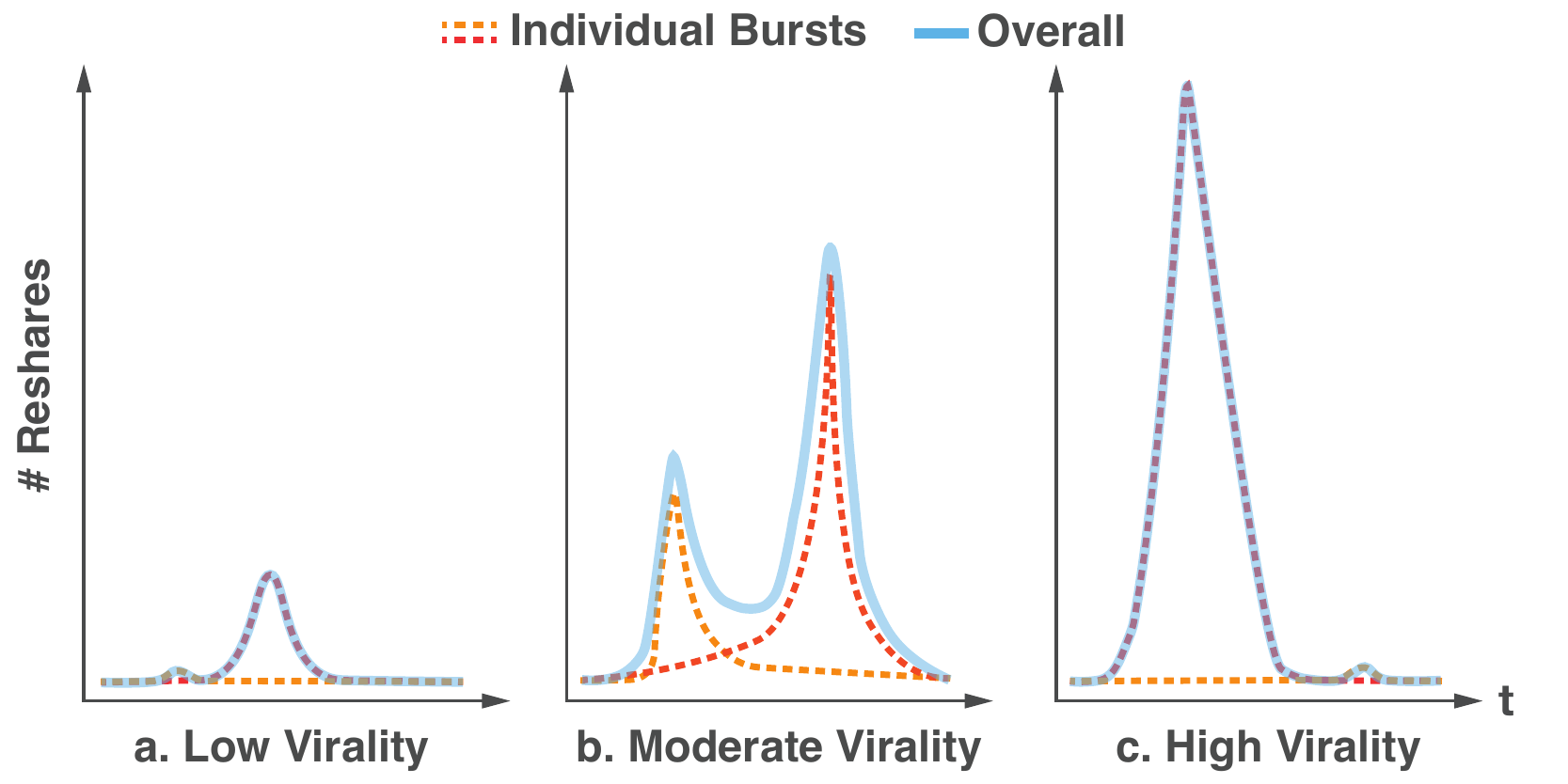}
    \caption{When virality is low, only a small number of attempts at infection succeed. When virality is moderate, more attempts succeed, which aggregate into observable recurrence. When virality is high, rather than a large number of bursts aggregating to form a single large peak, the first successful burst infects a large portion of the network, making it difficult for other copies to spread.}
    \label{fig:model}
\end{figure}

\subsection{A simple model of recurrence}

Motivated by these findings, we suggest a simple model of cascading behavior where recurrence depends on content virality:
\begin{itemize}[noitemsep]
\item If the virality of a cascade is low, it may only appeal to a small group of people, and is thus unable to spread far in the network. Thus, a single, small peak results, with many attempts to propagate in the network failing (Figure \ref{fig:model}a).
\item As virality increases, the cascade is able to spread substantially further in the network, and may occasionally even jump to other local communities in the network, spreading faster within them. As several bursts occur in the network, they may be observed as recurring in aggregate (Figure \ref{fig:model}b).
\item However, as virality increases beyond some threshold, any individual burst is likely to spread through a large portion of the susceptible population, inhibiting the transmission of subsequent copies (Figure \ref{fig:model}c). This last point lies in contrast to the trivial hypothesis that more independent copies leads to more independent bursts that aggregate to form a single large burst, which does not appear to be the case, as most reshares in initial bursts can be attributed to a single copy.
\end{itemize}

%% file: 045simulation.tex

\subsection{Simulating recurrence}
To see if such a model of recurrence can reproduce characteristics of recurrence observed in the data, we now simulate recurrence on a real social network.
Our observations and model suggest the use of an SIR model, where nodes in a network are initially susceptible (S) to a contagion, and then may become infected (I) when exposed. Infected nodes subsequently recover (R) and become resistant to the contagion. These models have been used to study the spread of disease \cite{bailey1975mathematical} and information \cite{bauckhage2015viral,cha2012delayed,newman2002spread} in a network.

\xhdr{Setup}
Our simulation thus consists of an SIR model with multiple outbreaks introduced at different times, and with resistant nodes reinfectable at a lower rate.
We parameterize our model as follows:
For a given contagion $c$, its virality, or equivalently, the susceptibility of every node in the network, is $p^c_0$.
In other words, if exposed to the contagion, the probability that the node will be infected is $p^c_0$.
Infected nodes attempt to infect all neighbors in the subsequent time step, and then become resistant.
As users sometimes share the same content multiple times, resistant nodes have a constant lower probability $p^c_1 < p^c_0$ of being re-infected.
The introduction of each copy of a contagion is normally distributed in time ($N(\mu, \sigma)$).

Here, we make a simplifying assumption that independent copies of the same content are introduced into the network at different points in time.
Following the intuition that more connected entities (e.g., pages) are likely to start outbreaks, the target nodes to infect are sampled, with replacement, proportional to the node's degree.
$m$ copies are introduced in total.

We simulate this model for 1000 discrete time steps with $\mu$=500, $\sigma$=250, and $m$=50, varying $p^c_0$ between 5\e{-4} and 10\textsuperscript{-3} and where $p^c_1$=$0.5 \cdot p^c_1$.
We run our simulation on the network of a country with approximately 1.4 million nodes and 160 million friendship edges, repeating the simulation 5000 times.
We measure the total number of infections (or reshares) in each time step, and identify bursts as defined in Section \ref{sec:definition}.

\begin{figure}[tb]
        \centering
        \begin{subfigure}[b]{0.49\columnwidth}
                \includegraphics[width=\textwidth]{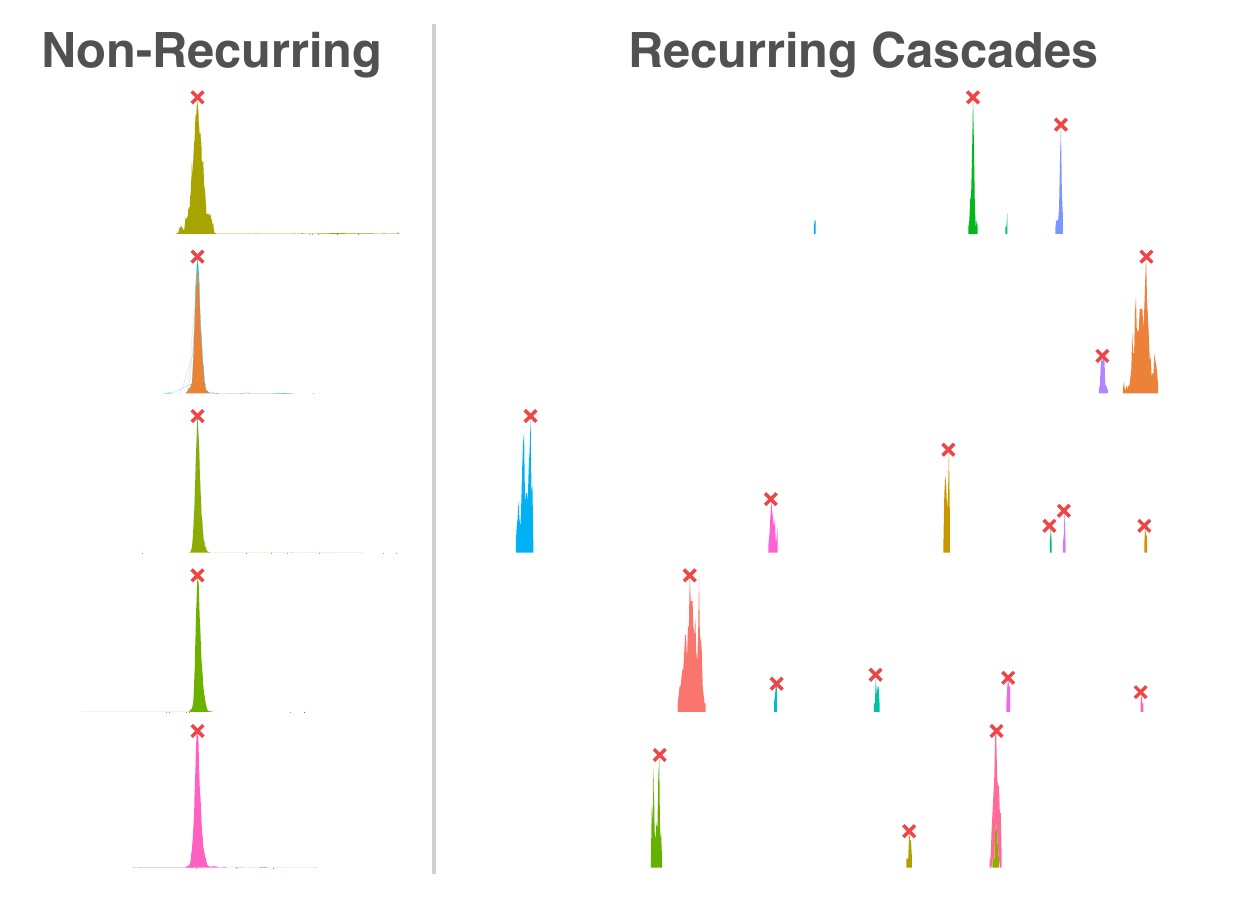}
                \caption{Simulated Time Series}
                \label{fig:model_examples}
        \end{subfigure}
        \begin{subfigure}[b]{0.49\columnwidth}
                \includegraphics[width=\textwidth]{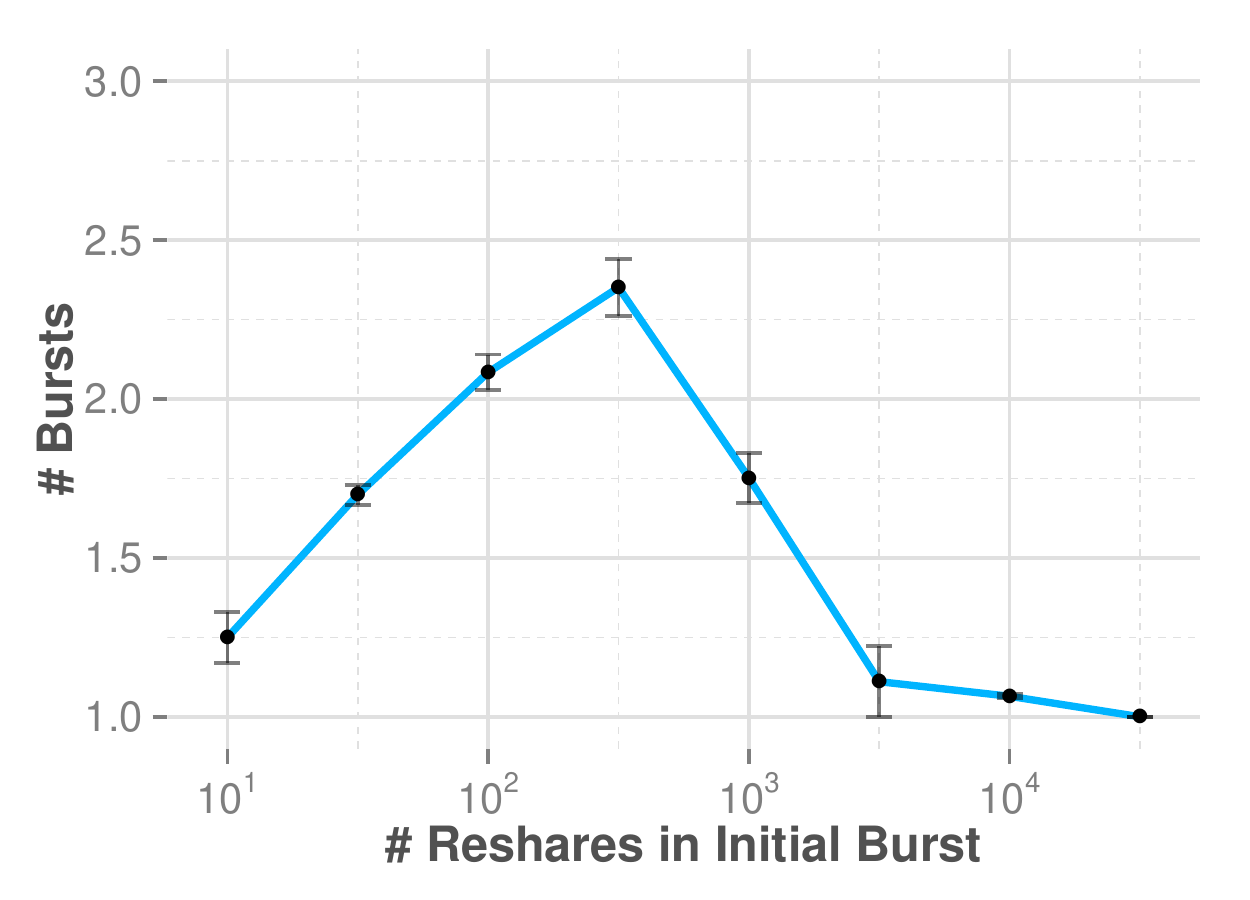}
                \caption{\# Peaks vs. Reshares, Simulated}
                \label{fig:model_imax}
        \end{subfigure}
        \caption{(a) By varying content virality, a model of recurrence that assumes independent introductions of copies of the same content can simulate recurrence. (b) It also replicates the observation that a moderate number of reshares results in more recurrence.}\label{fig:simulation}
\end{figure}

\xhdr{Results}
Within certain ranges of virality (6\e{-4}$\le p^c_0<$8\e{-4}), we can consistently reproduce recurring cascades.
Figure \ref{fig:model_examples} shows several examples of the time series of these simulations.
In aggregate, we can obtain a distribution of number of peaks similar in shape to Figure \ref{fig:peaks}.
Plotting the number of peaks against number of reshares in the initial burst (or alternatively, $p^c_0$), we observe an interior maximum --- a moderate amount of virality results in the most recurrence (Figure \ref{fig:model_imax}), replicating our previous findings.

When the virality of the contagion is high ($p^c_0$$\ge$8\e{-4}), a large fraction of the highly connected portion of the graph becomes infected by a single copy in the initial burst, suppressing subsequent bursts as many nodes are now resistant.
To show this happening, we consider for each simulation, in addition to our original model, an alternate-universe setting where the resistances of nodes are reset following the initial burst.
We can then measure how much the initial burst inhibited the second by observing the likelihood of a second burst in the alternate case, as well as the overlap of nodes infected in the second burst with nodes in the initial burst.
A significant difference in the total number of peaks when virality is high (1.0 vs. 2.0, \textit{t}=92, \textit{p}\textless10\textsuperscript{-10}), but not when virality is low ($p^c_0$$\le$7\e{-4}, n.s.) suggests that the supply of susceptible nodes is indeed being used up in the former case, but not the latter.
A significant positive correlation of initial peak size with the size of the overlap of the second peak in the alternate setting (0.76) further supports this hypothesis and our prior observations.

Likewise, the connectivity of graph deteriorates significantly after a large initial burst ($p^c_0$$\ge$8\e{-4}).
Here, we measure the algebraic connectivity \cite{fiedler1973algebraic} of the graph if all the nodes involved in the initial burst are removed, and compare this to a baseline that removes the same number of nodes at random.
Connectivity is significantly lower in the former case (579 vs. 1065, \textit{t}>17, \textit{p}\textless10\textsuperscript{-10}), especially in comparison to the graph's initial connectivity (1105).

These results together suggest that under such a model of recurrence, a large initial burst does indeed inhibit subsequent bursts, as we previously hypothesized (Figure \ref{fig:model}c).
Also in support of our prior observations, increasing the number of introduced copies $m$ monotonically increases recurrence.

\xhdr{Limitations and alternatives}
Importantly, our model assumes that recurrence is sparked primarily by independent copies introduced to the network.
However, the reality of recurrence is subtler: individual copies recur significantly in the network, and homophily may also moderate recurrence.
Allowing virality to vary with time \cite{guille2012predictive} or having nodes wait according to a power-law distribution \cite{crane2008robust,liben2008tracing} may also reproduce recurrence with only a single copy.
Decision-based queuing processes \cite{barabasi2005origin} may also help model the long periods of inactivity between bursts.

%% file: 050prediction.tex

Is it possible to predict if a cascade will resurface in the future?
Observing just the initial burst of a cascade, we use features related to the temporality, network structure, user demographics, and presence of multiple copies to determine
\begin{enumerate*}[label=\itshape\alph*\upshape)]
\item \emph{whether} recurrence occurs,
\item if the recurrence will be relatively \emph{smaller or larger}, and
\item \emph{when} the recurrence occurs
\end{enumerate*}.
Overall, we find that cascades with longer initial bursts that consist of multiple small outbreaks tend to recur, supporting the hypothesis that content virality and multiple copies play a significant role in recurrence.
Nonetheless, we obtain similarly strong performance predicting recurrence for individual copies of content.
Predicting recurrence may enable us to better forecast content longevity in a network.

\subsection{Factors driving recurrence}
Based on our observations, we develop several features that help predict recurrence, and group them into four categories:

\xhdr{Temporal features (7)}
Initially longer-lived bursts are suggestive of recurrence, motivating the importance of the \emph{number of days before} and \emph{after} the peak is reached, as well as the \emph{number of reshares before} and \emph{after}, and the \emph{height of the initial peak}.
The average \emph{gradient of the initial burst before} and \emph{after} the peak further characterize the shape of the initial burst.

\xhdr{Demographic features (5)}
The differences in user characteristics and diversity we previously observed suggest the importance of \emph{age}, \emph{gender}, as well as the \emph{entropy in the distribution of age}, \emph{gender} and \emph{country} of the initial burst.

\xhdr{Network features (6)}
Recurring cascades appear to be more connected in their initial bursts, having more \emph{friendship} and \emph{follower edges}, in addition to having a larger potentially \emph{exposed population}.
The \emph{number of users}, \emph{pages}, and \emph{proportion of pages} in the initial burst also vary.

\xhdr{Multiple-copy features (8)}
The availability of multiple copies plays a significant role in recurrence, motivating the use of the \emph{number of copies observed in the initial peak}, the \emph{entropy in the distribution of reshares of each copy}, the \emph{mean reshares per copy}, and the \emph{proportion of reshares attributable to the most popular copy}.
Pages also play a role in recurrence, suggesting that the \emph{proportion of copies created by pages}, the \emph{proportion of all reshares made by pages} or \emph{attributable to page-created copies}, and \emph{whether the most popular copy was created by a page} are useful features.

\begin{table}[tb]
\small
\centering
\ra{1.3}
\begin{tabular*}{\columnwidth}{@{\extracolsep{\fill}}llll}\toprule
	\raisebox{1.3ex}{AUC on Feature Sets} & \IconIfTable & \IconSizeTable & \IconWhenTable \\
	\hline
	Temporal & 0.74 & 0.76 & 0.55 \\
	+ Demographic & 0.78 \deem{(0.63)} & 0.76 \deem{(0.58)} & 0.56 \deem{(0.52)} \\
	+ Network & 0.81 \deem{(0.72)} & 0.77 \deem{(0.66)} & 0.57 \deem{(0.53)} \\
	+ Multiple-Copy & 0.89 \deem{(0.82)} & 0.78 \deem{(0.70)} & 0.58 \deem{(0.54)} \\
\bottomrule
\end{tabular*}
\caption{We obtain strong performance in predicting \emph{whether} recurrence occurs and if the subsequent burst will be \emph{smaller or larger}, but not in predicting \emph{when} recurrence occurs. Individual feature set performance is in parentheses. The column headers refer to Sections \ref{ssec:predict_if}, \ref{ssec:predict_size}, and \ref{ssec:predict_time} respectively.}
\label{tab:performance}
\end{table}

\subsection{Does it recur? \IconIf}
\label{ssec:predict_if}

\xhdr{Prediction task}
We formulate our prediction task as a binary classification problem: given only the initial burst of a cascade, we aim to predict if a second burst will be observed (i.e., if the cascade will recur).
We use a balanced dataset of recurring and non-recurring cascades (\textit{N}=40,912 for image memes, 89,368 for videos) so that guessing results in a baseline accuracy of 0.5.
Given the non-linear relation of several features to recurrence (e.g., that a moderate number of reshares results in the most recurrence), we use a random forest classifier.
In all cases, we perform 10-fold cross-validation and report the classification accuracy, F1 score, and area under the ROC curve (AUC).

\xhdr{Results}
Overall, we find strong performance in predicting recurrence (Accuracy=0.82, F1=0.81, AUC=0.89).
A logistic regression classifier results in slightly worse performance (AUC=0.78).
Table \ref{tab:performance} shows how performance improves as features are added to the model, as well as individual feature set performance.
While multiple-copy features perform best, temporal and network features, and to a lesser extent demographic features, also individually exhibit robust performance, suggesting that each significantly contributes to recurrence.
In the absence of strong multiple-copy features (fewer copies of any one video exist), we obtain worse performance in predicting the recurrence of videos (Acc=0.69, F1=0.66, AUC=0.76), with temporal features instead performing best.

For image meme cascades, the most predictive features of recurrence relate to cascades having multiple small outbreaks (fewer reshares per copy (0.78) and a higher entropy in the distribution of reshares across copies (0.72)), and longer initial bursts (more days before (0.63) and after (0.63) the peak).
These features remain important for video cascades.
Mirroring the dual importance of multiple-copy and temporal features, just the number of reshares per copy and the average gradient of the initial burst after its peak alone achieve strong performance (0.81).
Though the initial burst of a recurring cascade is on average significantly larger, size-related features are weaker signals of recurrence ($\le$0.59).

\subsection{Will the recurrence be smaller/larger? \IconSize}
\label{ssec:predict_size}

\xhdr{Prediction task}
Assuming that we know that a cascade will recur, how much smaller or larger will the second burst be?
Knowing the relative size of the next recurrence can differentiate bursty cascades that are rising or falling in popularity. 
Given the initial burst of a cascade, we aim to predict if the relative size of the second burst, or the ratio of the size of the second burst to that of the first, is above or below the median (0.28).
As the median evenly divides the dataset, we again have a balanced binary classification task with a random guessing baseline accuracy of 0.5.

\xhdr{Results}
We also find strong performance in predicting the relative size of the subsequent burst (Acc=0.72, F1=0.69, AUC=0.78 for image memes, AUC=0.85 for videos).
Temporal features here outperform all other feature sets, with the most predictive features relating to the cascade having a long initial burst.

\subsection{When does it recur? \IconWhen}
\label{ssec:predict_time}

\xhdr{Prediction task}
If a cascade will recur, when will we observe the next burst?
With a cascade's initial burst, can we predict if the duration between bursts will be greater than the median (14 days)?

\xhdr{Results}
We find that the timing of recurrence is far less predictable (Acc=0.56, F1=0.51, AUC=0.58 for image memes, AUC=0.60 for videos).
Nevertheless, longer initial bursts are most indicative of recurrence happening earlier.

\subsection{Predicting recurrence for individual copies}
Given the correlation of the appearance of multiple copies with bursts, multiple-copy features perform strongest in predicting recurrence.
But what if we want to predict recurrence of a single instance of some content, where multiple copies do not exist by definition?
Surprisingly, we obtain similarly strong performance in predicting the recurrence of individual copies (\textit{N}=28,454, Acc=0.80, F1=0.79, AUC=0.88 for image memes, AUC=0.82 for videos).
Network features are strongest (AUC=0.84), with fewer edges between users and pages (0.68) in the initial peak the most predictive of recurrence.
As individual copies have a single point of origin, fewer edges between pages and users and more edges between users (0.61) suggests that the burst may have resulted more from users sharing content from other users than high-degree pages sharing that content with their followers.
This observation, together with the fact that longer initial bursts continue to be strongly predictive of recurrence (\textgreater0.65), suggests the continued significance of virality with respect to individual copies.

The relative size of the subsequent burst is similarly predictable for individual copies (0.83 for image memes, 0.84 for videos), but interestingly, the time of recurrence is more predictable (0.68 and 0.63 respectively), which may be because any recurrence must be a continuation of the initial copy, as opposed to possibly being sparked by a new, less related copy.

%% file: 060related.tex

Significant prior work has studied information diffusion in online social media \cite{bakshy2012role,cheng2014can,myers2014bursty} --- with respect to memes, work has demonstrated the effect of meme similarity \cite{coscia2014average} and competition for limited attention on subsequent popularity \cite{weng2012competition}.
Most relevant is previous work that looked at the temporal dynamics of diffusion and developed epidemiological models of recurrence.

Among work that aims to predict the future popularity of online content \cite{borghol2011characterizing,lakkaraju2013s,stoddard2015popularity}, one relevant line of research has involved modeling the temporal patterns of the diffusion of information in social media \cite{ahmed2013peek,matsubara2012rise,yang2011patterns} or using these patterns to predict future popularity or forecast trends \cite{asur2010predicting,bauckhage2015viral,bauckhage2013mathematical,cheng2014can,choi2012predicting,guille2012predictive}.
Perhaps driven by the general shape of the initial burst of a cascade, many of these models implicitly assume that the temporal shape of a cascade consists primarily of a rising and falling period, and focus on modeling the initial activity around a peak \cite{ahmed2013peek,matsubara2012rise,yang2010modeling} or the overall popularity discounting subsequent spikes \cite{bauckhage2013mathematical}.
Beyond the initial burst of activity, we studied the long-term temporal dynamics of content on Facebook over a year.

In prior work, when multiple bursts are observed in a time series, they tend to be of a topic or hashtag rather than an individual piece of content, and are commonly attributed to external stimuli \cite{gruhl2004information,kumar2005bursty,leskovec2009meme,myers2012information} (e.g., news related to that topic).
While knowing about external events can help forecast the temporal pattern of the resulting spike \cite{matsubara2012rise}, there has been little work in predicting if new spikes will appear in the future lacking such knowledge.
In particular, rumor recurrence is bursty, with or without external stimuli, and sometimes with embellishments and other mutations \cite{adamic2014information,kwon2013prominent,friggeri2014rumor}, but there is little understanding of this phenomenon.
Patterns of human activity can also explain periodicity in popularity \cite{asur2011trends,grinberg2013extracting,leskovec2007patterns}, but the vast majority of recurrence we observe in this paper is aperiodic.
While external stimuli explains some instances of recurrence, we discover other factors that influence recurrence.
In contrast to most work that has observed multiple bursts in topics, we observed recurrence even at the level of an individual copy.

Finally, substantial work has studied how bursts in streams or time series can be detected \cite{kifer2004detecting,kleinberg2003bursty,palshikar2009simple}.
In this paper, we adopted a simple definition of burstiness, parameterizing peaks and bursts relative to the mean activity observed.

Recurrence has also been studied in the context of epidemiology, though primarily from a modeling perspective.
Many base their analysis on SIR models \cite{newman2002spread}, simulating recurrence through introducing dormant periods \cite{johansen1996simple}, seasonality effects \cite{altizer2006seasonality}, or changes in contagion fitness \cite{girvan2002simple}, which may be periodic \cite{olsen1988oscillations}.
More recently, some work studied content popularity using these models, while accounting for user login dynamics and content aging \cite{cha2012delayed}.
The structure of the network can also cause periodicity in epidemics \cite{kuperman2001small,verdasca2005recurrent}.
Many focus on modeling specific types of recurrence (e.g., historical disease epidemics \cite{altizer2006seasonality}).
In contrast, many recurrences we observe are aperiodic, and findings on synthetic networks may not easily generalize.
Inspired by this line of work, we adapted an SIR model assuming multiple points of infection on a real social network, and show that key characteristics of recurrence we observed can be reproduced.

%% file: 070conclusion.tex

Our results start to shed light on the mechanism of content recurrence --- studying a large dataset of popularly reshared content, we find that recurrence is common, and that content can come back not just once, but several times.
Strikingly, content may nearly cease to circulate for days, weeks or even months, prior to experiencing another surge in popularity.
Such a phenomenon may seem highly unpredictable, but we find trends in how recurring cascades behave, and can predict whether content will come back.
The virality, or appeal of a cascade plays a role in recurrence: cascades whose initial bursts are long-lasting, moderately popular, and moderately diverse are most likely to recur.
The presence of multiple copies of the same content sparks recurrence, though homophily in the network may also influence recurrence.

One limitation of our work is that we only analyze content within a single network.
Though most copies of the same content were made within the network, a minority appeared without a prior path.
Analyzing the transfer of content between different social networks may reveal different mechanisms of recurrence.
Separately, while the appearance of multiple copies correlates with recurrence, this does not hold in the case of individual-copy recurrence.
Understanding recurrence in the absence of multiple copies (e.g., through studying homophily in more detail) remains future work.

Based on our observations, we presented a simple model that exhibits some features of recurrence (e.g., pronounced bursts with little activity in-between, and an internal maximum in the number of bursts as a function of the number of reshares).
Future work could extend such models to account for homophily and community structure in the network.

While the temporal shape, network structure, and user attributes are already highly predictive of resharing behavior, other factors may improve prediction accuracy further: sentimentality or humor may make content evergreen, while content tied to current events may have an expiration date.
Seasonality effects may also cause periodic recurrence: we did observe an instance of a daylight-savings image meme which appeared, as expected, exactly at the two points during the year when people needed to adjust their clocks.
Also, other types of content may exhibit different properties of recurrence (e.g., link sharing may be more externally driven);
the interactions of users with shared content (e.g., comments) may also reveal the reasons why some content came back;
the societal context of memes, as well as their interactions (or competition) with other content, may also reveal more insight into their popularity \cite{spitzberg2014toward}.
Perhaps most suggestive that much remains to be studied is that while we can predict if recurrence will happens, it remains a significant challenge to predict \emph{when} recurrence will happen.

\xhdr{Acknowledgments} This work was supported in part by a Microsoft Research PhD Fellowship, a Simons Investigator Award, NSF Grants CNS-1010921 and IIS-1149837, and the Stanford Data Science Initiative.